\documentclass[12pt]{article}
\usepackage{graphics,color}
\usepackage{latexsym,amssymb}
\usepackage{amsmath,amsbsy}
\usepackage[all]{xy}
\usepackage{amsopn,amstext}
\usepackage{cite}
\usepackage{amssymb}
\usepackage{rotating}
\usepackage{amsmath}
\usepackage{amsopn,amstext}
\usepackage{epstopdf}
\allowdisplaybreaks[4]

\begin{document}

\def\bea{\begin{eqnarray}}
\def\eea{\end{eqnarray}}
\def\no{\nonumber}

\baselineskip=20pt

\newcommand{\Title}[1]{{\baselineskip=26pt
   \begin{center} \Large \bf #1 \\ \ \\ \end{center}}}
\newcommand{\Author}{\begin{center}
   \large \bf
Guang-Liang Li${}^{a,b}$, Yi Qiao${}^{c}\footnote{Corresponding author: qiaoyi\_joy@foxmail.com}$, Junpeng Cao${}^{b,c,d,e}\footnote{Corresponding author: junpengcao@iphy.ac.cn}$, Wen-Li
Yang${}^{b,f,g}\footnote{Corresponding author:
wlyang@nwu.edu.cn}$, Kangjie Shi${}^{f,g}$ and Yupeng
Wang${}^{b,c,h}$
 \end{center}}

\newcommand{\Address}{\begin{center}
${}^a$ Ministry of Education Key Laboratory for Nonequilibrium
Synthesis and Modulation of Condensed Matter, School of
Physics, Xi'an Jiaotong University, Xi'an 710049, China\\
${}^b$ Peng Huanwu Center for Fundamental Theory, Xi'an 710127, China\\
${}^c$ Beijing National Laboratory for Condensed Matter Physics, Institute of Physics, Chinese Academy of Sciences, Beijing 100190, China\\
${}^d$ School of Physical Sciences, University of Chinese Academy of Sciences, Beijing 100049, China\\
${}^e$ Songshan Lake Materials Laboratory, Dongguan, Guangdong 523808, China \\
${}^f$ Institute of Modern Physics, Northwest University, Xi'an 710127, China\\
${}^g$ Shaanxi Key Laboratory for Theoretical Physics Frontiers, Xi'an 710127, China\\
${}^h$ The Yangtze River Delta Physics Research Center, Liyang,
Jiangsu, China
\end{center}}

\Title{Exact surface energy of the $D^{(1)}_2$ spin chain with generic non-diagonal boundary reflections}

\Author

\Address \vspace{0.3truecm}
\begin{abstract}

The exact solution of the $D^{(1)}_2$ quantum spin chain with generic non-diagonal boundary reflections is obtained.
It is found that the generating functional of conserved quantities of the system
can be factorized as the product of transfer matrices of two anisotropic $XXZ$ spin chains with open boundary conditions.
By using the factorization identities and the fusion technique, the
eigenvalues and the Bethe ansatz equations of the model are obtained.
The eigenvalues are also parameterized by the zero roots of the transfer matrix, and
the patterns of root distributions are obtained.
Based on them, ground states energy and the surface energies induced by the twisted boundary magnetic fields in the thermodynamic limit are obtained.
These results are checked by the numerical calculations.
The corresponding isotropic limit is also discussed.
The results given in this paper are the foundation to study the exact physical properties of high rank $D^{(1)}_{n}$ model by using the nested processes.

\vspace{0.5truecm} \noindent {\it PACS:} 75.10.Pq, 02.30.Ik,
71.10.Pm

\end{abstract}
\newpage

\section{Introduction}
\label{intro} \setcounter{equation}{0}

The $D^{(1)}_2$ spin chain is a typical one-dimensional quantum integrable system and has many applications in the high energy, topological and mathematical physics.
The corresponding exact solution is the foundation to exactly solve the high rank $D^{(1)}_{n}$ spin chain by the nested methods.
Each spin of the $D^{(1)}_2$ chain has four components thus the integrable $D^{(1)}_2$ model is characterized by the $16\times 16$ $R$-matrix \cite{13,14,15}, which is the solution of Yang-Baxter equation.
Staring from the $R$-matrix, the transfer matrix and conserved quantities including the Hamiltonian of the system with periodic boundary condition can be constructed by using the quantum inverse scattering method. Eigenvalues of the transfer matrix of the periodic $D^{(1)}_n$ model are obtained by using the analytical Bethe ansatz \cite{25-1,25} and then by the algebraic Bethe ansatz \cite{m1}.

Later, it was found that the $D^{(1)}_2$ spin chain with certain open boundary conditions can also be solved exactly, where the
boundary reflections are quantified by the reflection matrices which are the solutions of reflection equations \cite{5-5,5-6, Ana}.
For the open boundary case, the conserved quantities are generated by the transfer matrix consisted of the $R$-matrices and reflection matrices \cite{op1,op2,Mez91}.
If the boundary reflection matrices only have the diagonal elements, the particle numbers of each spin-component are conserved.
In this case, the quasi-vacuum (or the reference) state is easy to be constructed, and
the eigenvalues and Bethe-type eigenstates of the transfer matrix and Bethe ansatz equations can be obtained by the nested algebraic Bethe ansatz \cite{a1,a2,a3,a4,a5}.
Then based on the Bethe states, the correlation function, norm, form factor and other scalar products can be calculated.
The interesting thing is that how to diagonalize the transfer matrix if the reflection matrices are non-diagonal. Due to
the fact that the reflection matrices at the two ends can not be diagonalized simultaneously, the $U(1)$ symmetry of the system is broken.
Then the traditional nested algebraic Bethe ansatz does not work.

Recently, in an interesting work \cite{b2}, Robertson, Pawelkiewicz, Jacobsen and Saleur showed that the $R$-matrix of the $D^{(2)}_2$ model is related with the
antiferromagnetic Potts model and the staggered $XXZ$ spin chain \cite{4,5,6,7,8,9,10}. It opens a new way to diagonalize the integrable model beyond the
$A$-series Lie algebra. Based on this idea and using the algebraic Bethe ansatz, Nepomechie and Retore obtained the exact solution of
both the closed and open $D^{(2)}_2$ spin chains \cite{b5}.

In this paper, we study the exact solution of the $D^{(1)}_2$ spin chain with non-diagonal boundary reflections.
We find that the transfer matrix of the model can be factorized as the product of transfer matrices of two six-vertex models with generic integrable open boundary condition.
With the help of this factorization identity and using the fusion,
we obtain the eigenvalues expressed by the inhomogeneous $T-Q$ relation and the Bethe ansatz equations of the model.
In order to study the physical properties of the system, we also
use the $t-W$ scheme and obtain the patterns of zero roots distributions.
Based on them, we obtain the ground state energy and the surface energy induced by the non-diagonal boundary magnetic fields in the thermodynamic limit
and check these results numerically.
The results of both the anisotropic and isotropic $D^{(1)}_2$ spin chains are given.

This paper is organized as follows. In section 2, we give a brief description of the anisotropic $D^{(1)}_2$ spin chain with open boundaries.
The $R$-matrix, reflection matrices and generating functional of conserved quantities are introduced.
In section 3, we show that the transfer matrix can be factorized as the product of two $XXZ$ spin chains.
In section 4, by using the fusion technique, we calculate the exact solution of the system. The inhomogeneous $T-Q$ relation and related Bethe ansatz equations are given.
In section 5, the thermodynamic limit and the surface energies of the anisotropic $D^{(1)}_2$ spin chain are studied.
In section 6, we list the results of the isotropic $D^{(1)}_2$ spin chain. The summary of main results and some concluding remarks are presented in section 7.

\section{Anisotropic $D^{(1)}_2$ spin chain}
 \setcounter{equation}{0}

Each spin in the $D^{(1)}_2$ model has four components. The anisotropic spin-exchanging interaction is quantified by the
$4^2\times 4^2$ $R$-matrix \cite{13,14,15}
\bea
 R_{12}(u)=\left(\begin{array}{cccc|cccc|cccc|cccc}
    a&&& &&&& &&&& &&&& \\
    &b&& &g_1&&& &&&& &&&& \\
    &&b& &&&& &g_1&&& &&&& \\
    &&&e &&&d_1& &&d_1&& &c_1&&& \\
   \hline &{g}_2&& &b&&& &&&& &&&& \\
    &&& &&a&& &&&& &&&& \\
    &&&{d}_2 &&&e& &&{c}&& &d_1&&& \\
    &&& &&&&b &&&& &&g_1&& \\
   \hline &&{g}_2& &&&& &b&&& &&&& \\
    &&&{d}_2 &&&{c}& &&e&& &d_1&&& \\
    &&& &&&& &&&a& &&&& \\
    &&& &&&&  &&&&b &&&g_1& \\
   \hline &&&{c}_2 &&&{d}_2& &&{d_2}&& &e&&& \\
    &&& &&&&{g}_2 &&&& &&b&& \\
    &&& &&&& &&&&{g_2} &&&b& \\
    &&& &&&& &&&& &&&&a \\
           \end{array}\right),\label{RD-matrix}
\eea
where $u$ is the spectral parameter, the non-zero matrix elements are \bea
&&a=2\sinh^2(\frac u2-2\eta), \;\;\;
b=2\sinh\frac u2\sinh(\frac u2-2\eta), \;\;\; e=2\sinh^2\frac u2, \;\;\;  c=2\sinh^2(2\eta), \nonumber\\
&& g_1=-2e^{-\frac u2}\sinh(2\eta)\sinh(\frac u2-2\eta),\;\;\; g_2=e^ug_1, \;\;\; d_1=2e^{-\frac u2}\sinh\frac u2\sinh (2\eta), \no\\
&& d_2=e^ud_1,\;\;\; c_1=2e^{-u}\sinh^2 (2\eta),\;\;\; c_2=2e^{u}\sinh^2 (2\eta), \label{RD21-element} \eea
and $\eta$ is the crossing or anisotropic parameter.
The $R$-matrix \eqref{RD-matrix} is defined in the tensor space $V_1 \otimes V_2$, where $V_1$ and $V_2$ are two four-dimensional linear spaces, and has the properties
\begin{eqnarray}
{\rm Periodicity}&:&R_{12}(u+2i\pi)=R_{12}(u),\nonumber\\
{\rm PT-symmetry}&:&R^{t_{1}t_{2}}_{12}(u)=R_{12}(u),\nonumber\\
{\rm Unitarity}&:&R_{12}(u)R_{21}(-u)=\rho(u)=4\sinh^2(\frac u2-2\eta)\sinh^2(\frac u2+2\eta),\nonumber\\
{\rm Initial\,\, condition}&:&R_{12}(0)=\rho(0)^{\frac{1}{2}}P_{12},\nonumber\\
{\rm Crossing \,\,unitarity}&:&R_{12}(u)^{t_{1}}{M}_{1}R_{21}(-u+8\eta)^{t_{1}}{M}_{1}^{-1}=\rho(u-4\eta),\label{Crossing-Unitarity-1}
\end{eqnarray}
where ${P}_{12}$ is the permutation operator with the matrix
elements $[{P}_{12}]^{\alpha\gamma}_{\beta\delta}=\delta_{\alpha\delta}\delta_{\beta\gamma}$, $R_{21}(u)={P}_{12}R_{12}(u){ P}_{12}$, $t_1$ (or $t_2$)
denotes the transposition in the subspace $V_1$ (or $V_2$) and $M_1$ is the $4\times 4$ diagonal constant matrix defined in the space $V_1$ \bea
{M}_1=M=diag(e^{4\eta},1,1,e^{-4\eta}).\eea
Besides, the $R$-matrix \eqref{RD-matrix} satisfies the Yang-Baxter equation
\begin{eqnarray}
R_{12}(u-v)R_{13}(u)R_{23}(v)=R_{23}(v)R_{13}(u)R_{12}(u-v), \label{yab}
\end{eqnarray}
which means that the scattering processes among the quasi-particle do not depend on the paths.

The boundary reflection at one end is characterized by the reflection matrix
\bea
K(u)=\left(\begin{array}{cccc}
    k_{11}(u)&k_{12}(u)&k_{13}(u)&k_{14}(u) \\
    k_{21}(u)&k_{22}(u)&k_{23}(u)&k_{24}(u) \\
    k_{31}(u)&k_{32}(u)&k_{33}(u)&k_{34}(u) \\
    k_{41}(u)&k_{42}(u)&k_{43}(u)&k_{44}(u)
\end{array}\right),\label{kd-2mat11rix}
\eea
where matrix elements are
\bea && k_{11}(u)=e^{-u}\sinh(s-\frac u2)\sinh(t-\frac u2),\quad k_{12}(u)=-e^{-\frac u2}t_1\sinh(s-\frac u2)\sinh u,\no \\
&& k_{13}(u)=e^{-\frac u2}s_1\sinh(t-\frac u2)\sinh u,\quad k_{14}(u)=s_1t_1\sinh^2 u,\no\\
&& k_{21}(u)=-e^{-\frac u2}t_2\sinh(s-\frac u2)\sinh u,\quad k_{22}(u)=\sinh(s-\frac u2)\sinh(t+\frac u2),\no\\
&& k_{23}(u)=-s_1t_2\sinh^2 u,\quad k_{24}(u)=-e^{\frac u2}s_1\sinh(t+\frac u2)\sinh u,\no\\
&& k_{31}(u)=e^{-\frac u2}s_2\sinh(t-\frac u2)\sinh u,\quad  k_{32}(u)=-s_2t_1\sinh^2 u,\no\\
&& k_{33}(u)=\sinh(s+\frac u2)\sinh(t-\frac u2),\quad  k_{34}(u)=e^{\frac u2}t_1\sinh(s+\frac u2)\sinh u,\no\\
&& k_{41}(u)=s_2t_2\sinh^2 u,\quad  k_{42}(u)=-e^{\frac u2}s_2\sinh(t+\frac u2)\sinh u,\no\\
&& k_{43}(u)=e^{\frac u2}t_2\sinh(s+\frac u2)\sinh u,\quad k_{44}(u)=e^{ u}\sinh(s+\frac u2)\sinh(t+\frac u2),
\eea
and $\{t, t_1, t_2, s, s_1, s_2\}$ are six free boundary parameters\footnote{The $K$-matrix (2.7) is a new K-matrix in the sense which has more non-vanishing matric elements
 and more boundary parameters than those of the $K$-matrix obtained by A. Lima-Santos \cite{5-5,5-6}.}. The reflection matrix \eqref{kd-2mat11rix} satisfies the reflection equation \cite{op1,op2}
\bea
R_{12}(u-v){K_{1}}(u)R_{21}(u+v) {K_{2}}(v)=
 {K_{2}}(v)R_{12}(u+v){K_{1}}(u)R_{21}(u-v),  \label{r1} \eea
where $K_{1}(u)=K(u)\otimes I$, $K_{2}(u)=I\otimes K(u)$ and $I$ is the $4\times 4$ unit matrix.
The boundary reflection at the other end is quantified by the dual reflection matrix
\bea
\bar {K}(u)=M K(-u+4\eta)|_{\{t, t_1, t_2, s, s_1, s_2\}\rightarrow \{t', t'_1, t'_2, s', s'_1, s'_2\}}, \label{kM}
\eea
where $\{t', t'_1,$ $ t'_2, s', s'_1, s'_2\}$ are the free boundary parameters.
The dual reflection matrix \eqref{kM} satisfies the dual reflection equation \cite{op1,op2,Mez91}
\begin{eqnarray}
&&R_{12}(-u+v){\bar {K}_{1}}(u)M_1^{-1}R_{21}
 (-u-v+8\eta)M_1{\bar {K}_{2}}(v)\nonumber\\[4pt]
&&\qquad={\bar {K}_{2}}(v)M_1R_{12}(-u-v+8\eta)M_1^{-1}
\bar K_{1}(u)R_{21}(-u+v), \label{r2}
\end{eqnarray}
where $\bar K_{1}(u)=\bar K(u)\otimes I$ and $\bar K_{2}(u)=I\otimes \bar K(u)$.

Now, we are ready to construct the quantum many-body system with interactions.
The conserved quantities including the model Hamiltonian of $D_2^{(1)}$ spin chain is generated by the transfer matrix $t(u)$
\begin{equation}
t(u)= tr_0 \{\bar K_0(u)T_0 (u) K_0(u)\hat{T}_0 (u)\}.
\label{trweweu1110}
\end{equation}
Here the subscript $0$ means the four-dimensional auxiliary space $V_0$ and
$tr_0$ means taking trace only in the auxiliary space.
$T_0(u)$ and $\hat{T}_0(u)$ are the monodromy matrix and the reflecting one, respectively,
\bea
&& T_0(u)=R_{01}(u-\theta_1)R_{02}(u-\theta_2)\cdots R_{0N}(u-\theta_N), \no\\
&& \hat{T}_0 (u)=R_{N0}(u+\theta_N)\cdots R_{20}(u+\theta_2) R_{10}(u+\theta_1),\label{Tt11}
\eea
where $\{\theta_j| j=1,\cdots, N\}$ are the inhomogeneity parameters and $N$ is the number of sites.
$T_0(u)$ and $\hat{T}_0(u)$ are defined in the tensor space $V_0 \otimes V_q$ and
$V_q=\otimes_{j=1}^N V_{j}$ is the physical space.
From the Yang-Baxter relation \eqref{yab}, reflection equation \eqref{r1} and the dual one \eqref{r2}, we can prove that
the transfer matrices with different spectral parameters commutate with each other
\begin{equation}
[t(u), t(v)]=0. \label{20211224-1}
\end{equation}
Thus the system is integrable. Expanding $t(u)$ with respect to
$u$, all the coefficients and their combinations are the conserved
quantities. The model Hamiltonian of the integrable $D^{(1)}_2$ quantum spin chain
is obtained by taking the derivative of logarithm of the transfer matrix $t(u)$ with the homogeneous limit $\{\theta_j\}=0$
\begin{eqnarray}
H&=&
\frac{\partial \ln t(u)}{2\partial u}|_{u=0,\{\theta_j\}=0}-\frac{tr_0{\bar{K}_0}(0)'}{2tr_0\bar{K}_0(0)} \nonumber \\
&=& \sum^{N-1}_{k=1}H_{k k+1}+\frac{{K_N}(0)'}{2{K_N}(0)}+\frac{
tr_0 \{ \bar{K}_0(0)H_{10}\}}{tr_0  \bar{K}_0(0)}, \label{hh}
\end{eqnarray}
where $H_{k\,k+1}= \rho^{-1}(0){
R}_{k\,k+1}(0)\,\frac{\partial}{\partial
u}R_{k\,k+1}(u)\left.\right|_{u=0}$ and
$H_{10}=P_{01}H_{01}P_{01}$.

The next task is to diagonalize the transfer matrix $t(u)$. However, the reflection matrices $K(u)$ and $\bar K(u)$
have the non-diagonal elements and can not be diagonalized simultaneously.
It is very hard to construct the reference state and to solve the eigen-equation of $t(u)$ \cite{wang15}.

\section{Factorizations}
\setcounter{equation}{0}

In order to obtain the eigenvalues of the transfer matrix \eqref{trweweu1110}, here we adopt the factorization method \cite{b2,b5}.
The four-dimensional space can be regarded as the tensor of two equivalent two-dimensional subspaces. For example, $V_1=V_{1'}\otimes V_{1''}$ and $V_2=V_{2'}\otimes V_{2''}$, where the structures
of $V_{1'}$ and $V_{1''}$ are the same. Then the $R$-matrix \eqref{RD-matrix} can be decomposed into
\bea
R_{12}(u)=2[S\otimes S][R^{s}_{1'2'}(u)\otimes R^{s}_{1''2''}(u)] [S\otimes S]^{-1},\label{Rs-matrissssx}\eea
where $S$ is a $4\times 4$ diagonal constant matrix
\bea S=diag(1,-1,1,1), \label{ai7}
\eea
and $R^s_{1'2'}(u)$ is the $R$-matrix of anisotropic $XXZ$ spin chain \bea
 R^s_{1'2'}(u)=\left(\begin{array}{cccc}
    \sinh(\frac u2-2\eta)&&&\\
    &\sinh\frac u2&-e^{-\frac u2}\sinh(2\eta)&  \\
    &-e^{\frac u2}\sinh(2\eta)&\sinh \frac u2&  \\
    &&&\sinh(\frac u2-2\eta) \\
           \end{array}\right).\label{Rs-matrix}
\eea
We shall note that the spaces of two $R$-matrices in the factorization \eqref{Rs-matrissssx} are different.
The $R$-matrix \eqref{Rs-matrix} has following properties
\begin{eqnarray}
{\rm Quasi-periodicity}&:&R^s_{1'2'}(u+2i\pi)=-R^s_{1'2'}(u),\nonumber\\
{\rm PT-symmetry}&:& R^s_{1'2'}(u)^{t_{1'}t_{2'}}=R^s_{1'2'}(u),\nonumber\\
{\rm Unitarity}&:&R^s_{1'2'}(u)R^s_{2'1'}(-u)=\rho_s(u)=\sinh(-\frac u2+2\eta)\sinh(\frac u2+2\eta),\nonumber\\
{\rm Initial\,\, condition}&:&R^s_{1'2'}(0)=-\sinh(2\eta) P_{1'2'},\nonumber\\
{\rm Crossing \,\,unitarity}&:&R^s_{1'2'}(u)^{t_{1'}}\bar{M}_{1'}R^s_{2'1'}(-u+8\eta)^{t_{1'}}\bar{M}_{1'}^{-1}=\rho_s(u-4\eta),\label{Crossing-Unita1rity-1}
\end{eqnarray}
where $P_{1'2'}$ is the $4\times 4$ permutation operator, $t_{1'}$ and $t_{2'}$ denote the transpositions in the subspaces $V_{1'}$ and $V_{2'}$, respectively,
and $\bar{M}_{1'}$ is a $2\times 2$ diagonal constant matrix $
\bar{M}_{1'}=\bar{M}=diag(e^{2\eta},e^{-2\eta})$. The $R$-matrix \eqref{Rs-matrix} satisfies Yang-Baxter equation
\bea
R^s_{1'2'}(u-v)R^s_{1'3'}(u)R^s_{2'3'}(v)= R^s_{2'3'}(v)R^s_{1'3'}(u)R^s_{1'2'}(u-v).
\eea

Following the same idea, the reflection matrices can be decomposed into
\bea
&&K(u)=S [K^{s+}(u)\otimes K^{s-}(u)] S^{-1}, \\[4pt]
&& \bar{K}(u)=S [\bar{K}^{s+}(u) \otimes \bar{K}^{s-}(u)]S^{-1},\label{po}\eea
where $K^{s\pm}(u)$ and $\bar K^{s\pm}(u)$ are the general reflection matrices of $XXZ$ spin chain
\bea
&& K^{s+}(u)=\left(\begin{array}{cc}
    -e^{-\frac u2}\sinh(\frac u2-s)&s_1\sinh u \\
   s_2\sinh u &e^{\frac u2}\sinh(\frac u2+s)
   \end{array}\right), \no \\
&&  K^{s-}(u)=K^{s+}(u)|_{\{s, s_1, s_2\}\rightarrow \{t, t_1, t_2\}}, \no \\
&&\bar K^{s+}(u)=\bar{M} K^{s+}(-u+4\eta)|_{\{s, s_1, s_2\}\rightarrow \{s', s'_1, s'_2\}}, \no \\
&& \bar K^{s-}(u)=\bar{M} K^{s-}(-u+4\eta)|_{\{t, t_1,
t_2\}\rightarrow \{t', t'_1, t'_2\}}, \label{kp-2} \eea which
satisfy the reflection equations \bea
\hspace{-0.8truecm}&&\hspace{-0.8truecm}
R^s_{1'2'}(u-v){K^{s\pm}_{  1'}}(u)R^s_{2'1'}(u+v)
{K^{s\pm}_{2'}}(v)=
 {K^{s\pm}_{2'}}(v)R^s_{1'2'}(u+v)K^{s\pm}_{1'}(u)R^s_{2'1'}(u-v), \no\\[4pt]
 \hspace{-0.8truecm}&&\hspace{-0.8truecm}R^s_{1'2'}(-u+v){\bar{K}^{s\pm}_{1'}}(u)\bar{M}_{1'}^{-1}R^{s\pm}_{2'1'}
 (-u-v+8\eta)\bar{M}_{1'}{\bar{K}^{s\pm}_{2'}}(v)\nonumber\\[4pt]
\hspace{-0.8truecm}&&\hspace{-0.8truecm}\qquad\qquad\qquad\qquad={\bar{K}^{s\pm}_{2'}}(v)\bar{M}_{1'}R^s_{1'2'}(-u-v+8\eta)\bar{M}_{1'}^{-1}
{{\bar{K}^{s\pm}}_{1'}}(u)R^s_{2'1'}(-u+v). \label{rs12}
\end{eqnarray}
Here ${K^{s\pm}_{1'}}(u)=K^{s\pm}(u)\otimes I'$,
${K^{s\pm}_{2'}}(u)=I'\otimes K^{s\pm}(u)$, $\bar K^{s\pm}_{  1'}(u)=\bar K^{s\pm}(u)\otimes I'$,
$\bar K^{s\pm}_{2'}(u)=I'\otimes \bar K^{s\pm}(u)$
and $I'$ is the $2\times 2$ unit matrix.

By using the factorizations \eqref{Rs-matrissssx} and \eqref{po}, we find that the transfer matrix
of $D^{(1)}_2$ spin chain can be factorized as the product of transfer matrices of two independent $XXZ$ spin chains
\bea t(u)=4^N{\cal S}\,t^{s+}(u)\otimes {t}^{s-}(u)\,{\cal S}^{-1}, \label{apo} \eea
where ${\cal{S}}=\otimes^N S$ and $t^{s\pm}(u)$ are defined as
\begin{equation}
t^{s\pm}(u)= tr_{0'} \{\bar K^{s\pm}_{0'}(u)T^s_{0'} (u) K^{s\pm}_{0'}(u)\hat{T}^s_{0'} (u)\}. \label{ts-1}
\end{equation}
We should note that the physical spaces of $t^{s+}(u)$ and $t^{s-}(u)$ in Eq.\eqref{ts-1} are different,
which means that they are two independent operators, and their tensor consists the physical space of the $D^{(1)}_2$ spin chain.
The monodromy matrix $T^s_{0'} $ and reflecting one $\hat{T}^s_{0'} (u)$
\bea
&& T^s_{0'}(u)=R^s_{0'1'}(u-\theta_1)R^s_{0'2'}(u-\theta_2)\cdots R^s_{0'N'}(u-\theta_N), \no\\
&& \hat{T}^s_{0'} (u)=R^s_{N'0'}(u+\theta_N)R^s_{N'-1\, 0'}(u+\theta_{N-1})\cdots R^s_{1'0'}(u+\theta_1), \label{Tt111}
\eea
satisfy the Yang-Baxter relations
\bea
&& R_{0'0''}^s(u-v) T^s_{0'}(u)  T^s_{0''}(v)= T^s_{0''}(v) T^s_{0'}(u) R^s_{0'0''}(u-v), \no\\
&& R_{0''0'}^s(u-v) \hat {T}^s_{0'}(u) \hat { T}^s_{0''}(v)=\hat { T}^s_{0''}(v)\hat { T}^s_{0''}(u) R^s_{0''0'}(u-v).\label{ai2-1}
\eea
By using the Yang-Baxter relation \eqref{ai2-1} and reflection equations \eqref{rs12}, we have
\bea
[t^{s+}(u), t^{s+}(v)]=0, \quad [t^{s-}(u), t^{s-}(v)]=0.
\eea

\section{Exact solution}
\setcounter{equation}{0}

The transfer matrices $t^{s+}(u)$ and $t^{s-}(u)$ can be
diagonalized separately. We first consider $t^{s+}(u)$. Following
the method developed in \cite{wang15}, we adopt  the fusion
technique \cite{f1,f2,f3,f4,f5,f6} to diagonalize the transfer
matrix $t(u)$ given by (\ref{trweweu1110}). The $R$-matrix
\eqref{Rs-matrix} at the point of $u=4\eta$ degenerates into \bea
R^s_{1'2'}(4\eta)=P^{(1) }_{1'2'}S_{1'2'}^{(1)}, \eea where
$S_{1'2'}^{(1)}$  is an irrelevant constant matrix omitted here,
$P^{(1)}_{1'2'}$ is the one-dimensional projector operator \bea
P^{(1)}_{1'2'}=|\psi_0\rangle\langle\psi_0|, \quad
|\psi_0\rangle=\frac{1}{\sqrt{2\cosh\eta}} (e^{-\eta
}|1'2'\rangle-e^{\eta }|2'1'\rangle), \eea and $\{|1'\rangle,
|2'\rangle\}$ are the orthogonal bases of $2$-dimensional linear
space $V_{1'}$ (and $V_{2'}$). Taking the fusion among the
$R$-matrices, we obtain \bea
&&P^{(1) }_{2'1'}R^s_{1'3'}(u)R^s_{2'3'}(u+4\eta)P^{(1)}_{2'1'}=\sinh(\frac u2+2\eta)\sinh(\frac u2-2\eta), \no \\[4pt]
&&P^{(1) }_{1'2'}R^s_{3'1'}(u)R^s_{3'2'}(u+4\eta)P^{(1)
}_{1'2'}=\sinh(\frac {u}{2}+2\eta)\sinh(\frac {u}{2}-2\eta).
\label{srf-2}
 \eea
The fusion of reflection matrices gives
\bea
\hspace{-1.0truecm}&&\hspace{-1.0truecm} P_{2'1'}^{(1)}K_{1'}^{s+}(u)R^s_{2'1'}(2u+4\eta)K_{2'}^{s+}(u+4\eta)P_{1'2'}^{(1)}\no\\[4pt]
\hspace{-1.0truecm}&&\hspace{-1.0truecm}\quad\quad\qquad\qquad=-2\sinh(u+4\eta)\frac{1}{\alpha}\cosh \frac{u+\alpha_1}{2}\cosh \frac{u-\alpha_1}{2}
\cosh\frac{u+\alpha_2}{2}\cosh\frac{u-\alpha_2}{2}, \no \\[4pt]
\hspace{-1.0truecm}&&\hspace{-1.0truecm}P_{1'2'}^{(1)}\bar{{K}}_{2'}^{s+}(u+4\eta)M_{1'}R^s_{1'2'}(-2u+4\eta)M_{1'}^{-1}
\bar{{K}}_{1'}^{s+}(u)P_{2'1'}^{\rm(1)}\no\\[4pt]
\hspace{-1.0truecm}&&\hspace{-1.0truecm}\quad\quad\qquad\qquad=2\sinh(u-4\eta)\frac{1}{\alpha'}\cosh \frac{u+\alpha'_1}{2} \cosh \frac{u-\alpha'_1}{2}
\cosh\frac{u+\alpha'_2}{2}\cosh\frac{u-\alpha'_2}{2}, \label{skf-2}
\eea
where the related constants are defined as \bea  &&\hspace{-1.0truecm}\alpha=\frac{1}{2s_1s_2},\quad
\cosh\alpha_1=\frac{\alpha}{2}+\beta, \quad \cosh\alpha_2=\frac{\alpha}{2}-\beta, \quad \beta=\sqrt{1+ \frac{\alpha^2}{4}+\alpha\cosh(2s)},\no\\[8pt]
&&\hspace{-1.0truecm}\alpha'=\frac{1}{2s'_1s'_2},\,\, \cosh\alpha'_1=\frac{\alpha'}{2}+\beta', \,\, \cosh\alpha'_2=\frac{\alpha'}{2}-\beta', \quad \beta'=\sqrt{1+\frac{\alpha'^2}{4}+\alpha'\cosh(2s')}.\label{alpha}
\eea
Yang-Baxter relations \eqref{ai2-1} at certain points gives
\bea
&&{T}^s_{0'}(\theta_j){T}_{0''}^s(\theta_j+4\eta)=P^{(1)}_{0''0'}{T}^s_{0'}(\theta_j){T}_{0''}^s(\theta_j+4\eta), \no\\[4pt]
&&\hat{T}^s_{0'}(-\theta_j)\hat{T}^s_{0''}(-\theta_j+4\eta)=P^{(1)}_{0'0''}\hat{T}^s_{0'}(-\theta_j)\hat{T}^s_{0''}(-\theta_j+4\eta).
\label{stht-1}
\eea

By using the fusion relations (\ref{srf-2})-(\ref{stht-1}), we obtain
\bea
\hspace{-0.8truecm}&&\hspace{-0.8truecm}
t^{s+}(\pm\theta_j) \,t^{s+}(\pm\theta_j+4\eta)=\frac{4\sinh(\pm{\theta_j}-4\eta)\sinh(\pm{\theta_j}+4\eta)}
{\alpha\alpha'\sinh(\pm\theta_j-2\eta)\sinh(\pm\theta_j+2\eta)}\cosh \frac{\pm\theta_j-\alpha_1}{2}\no\\[4pt]
\hspace{-0.8truecm}&& \hspace{-0.8truecm}\qquad\qquad \times
\cosh\frac{\pm\theta_j+\alpha_1}{2}
\cosh\frac{\pm\theta_j-\alpha_2}{2}\cosh\frac{\pm\theta_j+\alpha_2}{2}
\cosh\frac{\pm\theta_j-\alpha'_1}{2}\cosh\frac{\pm\theta_j+\alpha'_1}{2}
\no\\[4pt]
\hspace{-0.8truecm}&&\hspace{-0.8truecm}\qquad\qquad\times
\cosh \frac{\pm\theta_j-\alpha'_2}{2}\cosh\frac{\pm\theta_j+\alpha'_2}{2}
\prod_{k=1}^N\sinh\frac{\pm\theta_j-\theta_k-4\eta}{2}\no\\[4pt]
\hspace{-0.8truecm}&&\hspace{-0.8truecm}\qquad\qquad\times\sinh\frac{\pm\theta_j-\theta_k+4\eta}{2}
\sinh\frac{\pm\theta_j+\theta_k-4\eta}{2}\sinh\frac{\pm\theta_j+\theta_k+4\eta}{2}, \; j=1,\cdots, N. \label{sottf1}
\eea
Besides, the values of $t^{s+}(u)$ at the points of $u=0, 4\eta, i\pi$ can be calculated directly
\bea &&t^{s+}(0)=t^{s+}(4\eta)=2\cosh(2\eta)\sinh s\sinh s'\prod_{j=1}^N\rho_s(\theta_j),\\
&&t^{s+}(i\pi)=t^{s+}(-i\pi+4\eta)=2\cosh(2\eta)\cosh s\cosh s'\prod_{j=1}^N\rho_s(\theta_j+i\pi).\label{sottf3} \eea
The asymptotic behavior of  $t^{s+}(u)$ is
\bea t^{s+}(u)|_{u\rightarrow
\pm\infty}=-2^{-(2N+2)}e^{\pm[(N+2)u-2(N+2)\eta]}(e^{-2\eta}s_1s'_2+e^{2\eta}s_2s'_1). \label{sottf23}\eea

From the definition, we know that the transfer matrix $t^{s+}(u)$ is an operator polynomial of $e^{\frac u2}$ with the degree $2N+4$,
which can be completely determined by $2N+5$ constraints. Thus above $2N$
fusion identities \eqref{sottf1} and $6$ additional conditions \eqref{sottf3}-\eqref{sottf23} give us sufficient information to determine the eigenvalue $\Lambda^{s+}(u)$ of $t^{s+}(u)$.
After some algebra, we express the eigenvalue $\Lambda^{s+}(u)$ as the inhomogeneous $T-Q$ relation
\bea
\hspace{-0.8truecm}&&\hspace{-0.8truecm}\Lambda^{s+}(u)=\frac{2\sinh(u-4\eta)}
{\sinh(u-2\eta)\sqrt{\alpha\alpha'}}
\cosh\frac{u+\alpha_1}{2}\cosh\frac{u+\alpha_2}{2}\cosh\frac{u+\alpha'_1}{2}\cosh\frac{u+\alpha'_2}{2}\no\\[4pt]
\hspace{-0.8truecm}&&\hspace{-0.8truecm}\quad\quad\qquad\times
 \bar{a}(u)\frac{Q(u+4\eta)}{Q(u)}+\frac{2\sinh u}
{\sinh(u-2\eta)\sqrt{\alpha\alpha'}}\cosh\frac{u-4\eta-\alpha_1}{2}\cosh\frac{u-4\eta-\alpha_2}{2}\no\\[4pt]
\hspace{-0.8truecm}&&\hspace{-0.8truecm}\quad\quad\qquad\times
\cosh\frac{u-4\eta-\alpha'_1}{2}\cosh\frac{u-4\eta-\alpha'_2}{2}\bar{d}(u)\frac{Q(u-4\eta)}{Q(u)}\no\\[4pt]
\hspace{-0.8truecm}&&\hspace{-0.8truecm}\qquad\qquad +x_+\sinh u\sinh(u-4\eta)\frac{\bar{a}(u)\bar{d}(u)}{Q(u)}, \label{tse}\eea
where \bea &&Q(u)=\prod_{l=1}^{N}\sinh\frac 12(u-\mu_l)\sinh\frac
12(u+\mu_l-4\eta), \no\\[4pt]
&&\bar{a}(u)=\prod_{j=1}^N\sinh(\frac{u-\theta_j-4\eta}{2})
\sinh(\frac{u+\theta_j-4\eta}{2})=\bar{d}(u-4\eta),\no \\[4pt]
&&x_+=-2\sqrt{s_1s_2s'_1s'_2}\cosh[2(L+1)\eta+\frac{\alpha_1+\alpha_2+\alpha'_1+\alpha'_2}{2}]\no\\[4pt]
&& \quad \qquad -(e^{-2\eta}s_1s'_2+e^{2\eta}s_2s'_1).\no\eea
Because $\Lambda^{s+}(u)$ is a polynomial of $e^{\frac u2}$, the singularities of right hand side of Eq.\eqref{tse} should be cancelled with each other,
which give that the Bethe roots $\{\mu_l\}$ should satisfy the Bethe ansatz equations
\bea
\hspace{-0.8truecm}&&\hspace{-0.8truecm}\frac{2\sinh(\mu_l-4\eta)}
{\sinh(\mu_l-2\eta)\sqrt{\alpha\alpha'}}
\cosh\frac{\mu_l+\alpha_1}{2}\cosh\frac{\mu_l+\alpha_2}{2}\cosh\frac{\mu_l+\alpha'_1}{2}\cosh\frac{\mu_l+\alpha'_2}{2}
\frac{Q(\mu_l+4\eta)}{\bar{d}(\mu_l)}\no\\[4pt]
\hspace{-0.8truecm}&&\hspace{-0.8truecm}\quad\quad+\frac{2\sinh\mu_l}
{\sinh(\mu_l-2\eta)\sqrt{\alpha\alpha'}}\cosh\frac{\mu_l-4\eta-\alpha_1}{2}
\cosh\frac{\mu_l-4\eta-\alpha_2}{2}\cosh\frac{\mu_l-4\eta-\alpha'_1}{2}\no\\[4pt]
\hspace{-0.8truecm}&&\hspace{-0.8truecm}\quad\quad\times
\cosh\frac{\mu_l-4\eta-\alpha'_2}{2}
\frac{Q(\mu_l-4\eta)}{\bar{a}(\mu_l)}=-x_+\sinh \mu_l \sinh(\mu_l-4\eta),\;\; l=1,\cdots, N. \label{BA0} \eea

The eigenvalue $\Lambda^{s-}(u)$ of $t^{s-}(u)$ can be obtained by the mapping
\bea
\Lambda^{s-}(u)=\Lambda^{s+}(u)|_{\{s, s_1, s_2, s', s'_1, s'_2\}\rightarrow \{t, t_1, t_2, t', t'_1, t'_2\}}.\label{tse-1}
\eea
Substituting Eqs.\eqref{tse} and \eqref{tse-1} into the factorization identity \eqref{apo}, we obtain the eigenvalue $\Lambda(u)$ of transfer matrix $t(u)$
of the $D^{(1)}_2$ spin chain as
\bea\Lambda(u)=4^N\Lambda^{s+}(u)\Lambda^{s-}(u).\label{ta} \eea

Some remarks are in order. The solutions of algebraic equations
(\ref{BA0}) gives the values of Bethe roots $\{\mu_l\}$.
Substituting these values into the inhomogeneous $T-Q$ relation
(\ref{tse}), we obtain the eigenvalue $\Lambda^{s+}(u)$. As proven
in \cite{Cao14, Cao15}, the $T-Q$ relation (\ref{tse}) can
generate all the values of $\Lambda^{s+}(u)$. These results are
also valid for $\Lambda^{s-}(u)$. Then we conclude that the
expression (\ref{ta}) can give the complete spectrum of the
transfer matrix $t(u)$ of the $D^{(1)}_2$ spin chain. All the
above results have the well-defined homogeneous limit
$\{\theta_j\}=0$.

With the help of Eq.\eqref{hh}, we obtain the Hamiltonian of $D^{(1)}_2$ spin chain
\bea&&H=-\sum_{j=1}^{N-1}\frac{1}{4\sinh(2\eta)}[\cosh(
2\eta)(\sigma^z_j\sigma^z_{j+1}+\tau^z_j\tau^z_{j+1})+2(
\sigma^z_j\sigma^z_{j+1}(\tau^+_j\tau^-_{j+1}+\tau^-_j\tau^+_{j+1})\no\\
&&\qquad\quad +(\sigma^+_j\sigma^-_{j+1}
+\sigma^-_j\sigma^+_{j+1})\tau^z_j\tau^z_{j+1})]-\frac{1}{2\sinh
s\sinh t}(\frac12e^t\sinh s\tau^z_{N} +\frac12e^s\sinh
t\sigma^z_{N}\no\\[4pt]
&&\qquad\quad+t_1\sinh s\sigma^z_{N}\tau^+_{N}-s_1\sinh
t\sigma^+_{N}\tau^z_{N}+t_2\sinh s\sigma^z_{N}\tau^-_{N}-s_2\sinh
t\sigma^-_{N}\tau^z_{N})\no\\[4pt]
&&\qquad\quad+\frac{1}{2\sinh s'\sinh t'}(\frac12e^{t'}\sinh
s'\tau^z_{1}+\frac12e^{s'}\sinh t'\sigma^z_{1} +e^{2\eta}t'_1\sinh
s'\sigma^z_{1}\tau^+_{1}\no\\[4pt]
&&\qquad\quad-e^{2\eta}s'_1\sinh t'\sigma^+_{1}\tau^z_{1}
+e^{-2\eta}t'_2\sinh
s'\sigma^z_{1}\tau^-_{1}-e^{-2\eta}s'_2\sinh t'\sigma^-_{1}\tau^z_{1})\no\\[4pt]
&&\qquad\quad -\frac14(\sigma^z_{1}+\tau^z_{1}-\sigma^z_{N}-\tau^z_{N})-
\frac N2\coth 2\eta+\frac14\tanh2\eta\frac{\sinh (s'+t')}{\sinh
s'\sinh t'}, \label{apopd} \eea where
$\sigma_j^{\pm}=(\sigma_j^x\pm i \sigma_j^y)/2$,
$\tau_j^{\pm}=(\tau_j^x\pm i \tau_j^y)/2$, $\sigma_j^\alpha$ and
$\tau_j^\alpha$ are the Pauli matrices at $j$-th site and
$\alpha=x, y, z$. We should note that the Hamiltonian
\eqref{apopd} is the direct summation of two anisotropic $XXZ$
spin-1/2 chains with non-diagonal boundary magnetic fields
up to the similar transformation ${\cal S}$ \bea H={\cal S} [H^{s+}\oplus H^{s-}]{\cal
S}^{-1}, \label{apop1d}\eea where
\bea &&H^{s+}=-\sum_{j=1}^{N-1}
\frac{1}{4\sinh2\eta}[\sigma^x_j\sigma^x_{j+1}+\sigma^y_j\sigma^y_{j+1}+\cosh(2\eta)
\sigma^z_j\sigma^z_{j+1}]\no\\[4pt]
&&\qquad\quad+ \frac{1}{4\sinh s}[(s_1+s_2)\sigma^x_N +{i}(s_1-s_2)
\sigma^y_{N} -e^s
\sigma^z_N]+\frac14\sigma^z_{N}\no\\[4pt]
&&\qquad\quad-\frac{1}{4\sinh
s'}[(e^{2\eta}s'_1+e^{-2\eta}s'_2)\sigma^x_1
 +{i}(e^{2\eta}s'_1-e^{-2\eta}s'_2) \sigma^y_{1} -e^{s'} \sigma^z_1]-\frac14\sigma^z_{1}\no\\[4pt]
&&\qquad\quad-\frac{N}{4}\coth 2\eta
 +\frac14\tanh 2\eta\coth s', \label{H1}\\[6pt]
&&H^{s-}=H^{s+}|_{ \{\sigma_j^\alpha, s, s_1, s_2, s', s'_1,
s'_2\}\rightarrow \{\tau_j^\alpha, t, t_1, t_2, t', t'_1, t'_2\}},
\label{Hs}\eea with ${\cal{S}}=\otimes^N S$ and
$S=\frac{1}{2}(1-\sigma^z+\tau^z+\sigma^z\tau^z)$. The conclusion
\eqref{apop1d} is consistent with the fact that the corresponding
$R$-matrix and reflection matrices in the transfer matrix $t(u)$
of the $D_2^{(1)}$ spin chain can be factorized. The Hamiltonians
(\ref{H1}) and (\ref{Hs}) can also be generalized by the transfer
matrix $t^{s\pm}(u)$ as \bea H^{s\pm}=\frac{\partial \ln
t^{s\pm}(u)}{2\partial u}|_{u=0,\{\theta_j\}=0}+\frac14\tanh
2\eta(1+\coth c_{\pm}),\label{H1def} \eea where $c_+=s'$ and
$c_-=t'$.

The eigenvalue of the Hamiltonian of $D_2^{(1)}$ model
\eqref{apopd} is \bea E=\frac{\partial \ln \Lambda(u)}{2\partial
u}|_{u=0,\{\theta_j\}=0}+\frac14\tanh 2\eta(2+\coth s'+\coth t').
\label{taa} \eea In the derivation, we have used the useful
relations \bea
&&tr_0{\bar{K}_0}(0)'=-(\sinh(s'+t')+2\sinh s'\sinh t')\sinh 4\eta, \nonumber \\
&&tr_0{\bar{K}_0}(0)=4\cosh^22\eta\sinh s'\sinh t'.\nonumber \eea
Substituting the eigenvalue (\ref{ta}) into (\ref{taa}), we obtain
the eigen-energy of the $D_2^{(1)}$ spin chain.

\section{Thermodynamic limit}
\label{Thermo} \setcounter{equation}{0}

\subsection{$t-W$ scheme}

The Bethe roots in the energy spectrum \eqref{taa} should satisfy the Bethe ansatz equations \eqref{BA0}. However,
the present Bethe ansatz equations are inhomogeneous.
The corresponding patterns of Bethe roots are very complicated. Thus the analytical integral equations satisfied by the densities of Bethe roots in the thermodynamic limit are very hard to derive.
In order to overcome this difficulty, we use the $t-W$ scheme \cite{tm1,tm2}.

Due to the fact that the $D_2^{(1)}$ model can be factorized as two $XXZ$ spin-1/2 chains,
we consider the spin-1/2 chain first. Rewrite the Hamiltonian \eqref{H1} as
\bea &&H^{s+}=-\sum_{j=1}^{N-1}
\frac{1}{4\sinh2\eta}[\sigma^x_j\sigma^x_{j+1}+\sigma^y_j\sigma^y_{j+1}+\cosh(2\eta)
\sigma^z_j\sigma^z_{j+1}]+ h_1^+\sigma^+_1 +h_1^-\sigma^-_1\no\\[4pt]
&&\qquad+h_1^z\sigma^z_1
+ h_N^+\sigma^+_N +h_N^-\sigma^-_N+h_N^z\sigma^z_N-\frac{N}{4}\coth 2\eta
 +\frac14\tanh 2\eta\coth s',\label{Hs1}\eea
where the strengthes of boundary magnetic fields are quantified as
\bea
&&h_1^+=-\frac{e^{2\eta}s'_1}{2\sinh s'},\quad h_1^-=-\frac{e^{-2\eta}s'_2}{2\sinh s'},\quad h_1^z=\frac{\coth s'}{2}, \no \\
&&h_N^+=\frac{s_1}{2\sinh s},\quad h_N^-=\frac{s_2}{2\sinh s},\quad h_N^z=-\frac{\coth s}{2}.
\eea

In this paper, we consider the case that the anisotropic parameter $\eta>0$. The hermitian of the Hamiltonian \eqref{Hs1}
requires that $h_{1}^z$ and $h_N^z$ are real, $h_1^+=h_1^{-*}$ and $h_N^+=h_N^{-*}$, where the superscript $*$ means the complex conjugate.
According to the parametrization \eqref{alpha}, the hermitian requires that the boundary parameters $\alpha_1$ and $\alpha'_1$ are real, $\alpha_2=\bar{\alpha}_2+\pi i$ and $\alpha'_2=\bar{\alpha}'_2+\pi i$, where $\bar{\alpha}_2$ and $\bar{\alpha}'_2$ are real. These conclusions are obtained as follows.
The real $h_N^z$ gives that $s$ is real or $s=\bar{s}+\frac{i\pi}2$ where $\bar{s}$ is  real. For the case of real $s$, from Eq.\eqref{alpha} we have $\beta>1+\frac{\alpha_2}2$ and $\cosh\alpha_2<-1$. Thus we
denote $\alpha_2=\bar{\alpha}_2+\pi i$, where $\bar{\alpha}_2$ is real. For the case of $s=\bar{s}+\frac{i\pi}2$, we obtain the same results.
The constraints of boundary parameters $\alpha_1$, $\alpha'_1$ and $\alpha'_2$ can be obtained similarly.

With the help of Eq.\eqref{H1def}, the energy of the Hamiltonian \eqref{Hs1} is
\bea
E^{s+}=\frac{\partial \ln \Lambda^{s+}(u)}{2\partial u}|_{u=0,\{\theta_j\}=0}+\frac14\tanh
2\eta(1+\coth s'). \label{energy2exp}
\eea
Because the eigenvalue $\Lambda^{s+}(u)$ is a polynomial of $e^{\frac u2}$ with the degree $2N+4$, we parameterize it by the $2N+4$ zero points $\{z_k\}$ instead of the Bethe roots as
\bea
\Lambda^{s+}(u)=\Lambda_0\prod^{2N+4}_{k=1}\sinh\frac{u-z_k i-2\eta}{2},\label{Lamz}
\eea
where the coefficient $\Lambda_0=-4(e^{-2\eta}s_1s'_2+e^{2\eta}s_2s'_1)$, which can be calculated directly from the asymptotic behavior of transfer matrix $t^{s+}(u)$ with $u\rightarrow \infty$.
We note that the contributions of inhomogeneity parameters $\{\theta_j\}$ are included in the zero roots $\{z_k\}$.
Substituting Eq.\eqref{Lamz} into \eqref{energy2exp}, we have
\bea
E^{s+}=-\frac14\sum^{2N+4}_{k=1}\coth\frac{2\eta+\hat{z}_k i}2+\frac14{\tanh(2\eta)}(1+\coth s'), \label{energyexp}
\eea
where $\{\hat{z}_k\}$ are the homogeneous limit of zero roots $\{z_k\}$.

Now, we should determine the values of $2N+4$ zero roots $\{z_k\}$.
Acting the operator identities \eqref{sottf1}-\eqref{sottf3} on a common eigenstate $|\phi\rangle$ and considering the parametrization \eqref{Lamz}, we obtain
that the zero roots $\{z_k\}$ should satisfy the following $2N+4$ Bethe-ansatz-like equations
\bea
&&\Lambda_0^2\prod^{2N+4}_{k=1}\sinh\frac{\pm\theta_j- z_k i-2\eta}{2}\sinh\frac{\pm\theta_j- z_k i+2\eta}{2}=\frac{4\sinh(\pm\theta_j-4\eta)\sinh(\pm\theta_j+4\eta)}
{\alpha\alpha'\sinh(\pm\theta_j-2\eta)\sinh(\pm\theta_j+2\eta)}\no\\[4pt]
&&\times \cosh \frac{\pm\theta_j-\alpha_1}{2}
\cosh\frac{\pm\theta_j+\alpha_1}{2}
\cosh\frac{\pm\theta_j-\alpha_2}{2}\cosh\frac{\pm\theta_j+\alpha_2}{2}
\cosh\frac{\pm\theta_j-\alpha'_1}{2}
\no\\[4pt]
&&\times \cosh\frac{\pm\theta_j+\alpha'_1}{2}
\cosh \frac{\pm\theta_j-\alpha'_2}{2}\cosh\frac{\pm\theta_j+\alpha'_2}{2}
\prod_{k=1}^N\sinh\frac{\pm\theta_j-\theta_k-4\eta}{2} \no\\[4pt]
&&\times \sinh\frac{\pm\theta_j-\theta_k+4\eta}{2}
\sinh\frac{\pm\theta_j+\theta_k-4\eta}{2}\sinh\frac{\pm\theta_j+\theta_k+4\eta}{2}, \quad j=1,\cdots, N, \label{sollf1}
\eea
and
\bea &&\Lambda_0\prod^{2N+4}_{k=1}\sinh\frac{z_k i\pm 2\eta}{2}=2\cosh(2\eta)\sinh s\sinh s'\prod_{j=1}^N\rho_s(\theta_j),\label{t0}\\
&&\Lambda_0(-1)^N\prod^{2N+4}_{k=1}\cosh\frac{z_k i\pm 2\eta}{2}=2\cosh(2\eta)\cosh s\cosh s'\prod_{j=1}^N\rho_s(\theta_j+i\pi).\label{tipi} \eea

By solving the Bethe-ansatz-like equations \eqref{sollf1}-\eqref{tipi}, we can obtain the values of zero roots $\{z_k\}$ with the given inhomogeneity parameters $\{\theta_j\}$.
Taking the homogeneous limit and substituting the results into Eq.\eqref{energyexp}, we obtain the eigen-energy of the Hamiltonian \eqref{Hs1}.

\subsection{Patterns of zero points}

Now, we should determine the solutions of Bethe-ansatz-like equations \eqref{sollf1}-\eqref{tipi}.
From the definition of transfer matrix $t^{s+}(u)$ and using the crossing unitarity \eqref{Crossing-Unita1rity-1},
we find that the $t^{s+}(u)$ and its eigenvalue $\Lambda^{s+}(u)$ satisfy the crossing symmetry
\bea
t^{s+}(u)=t^{s+}(-u+4\eta), \quad \Lambda^{s+}(u)=\Lambda^{s+}(-u+4\eta).\label{crosssym}
\eea
Substituting \eqref{Lamz} into \eqref{crosssym}, we have
\bea
\prod^{2N+4}_{k=1}\sinh\frac{u-z_k i-2\eta}{2}=\prod^{2N+4}_{k=1}\sinh\frac{u+z_k i-2\eta}{2}.
\eea
Thus we conclude if $z_k$ is a zero root of $\Lambda^{s+}(u)$, then $-z_k$ must be the root.

If the inhomogeneity parameters $\{\theta_k\}$ are pure imaginary or zero, by using the intrinsic properties of $R$-matrix \eqref{Rs-matrix},
one can easily prove that
\bea\label{lambda-conj}
[t^{s+}(u)]^\dagger=t^{s+}(u^*), \quad [\Lambda^{s+}(u)]^*=\Lambda^{s+}(u^*).
\eea
Substituting \eqref{Lamz} into \eqref{lambda-conj}, we have
\bea
\prod^{2N+4}_{k=1}\sinh\frac{u^*+z_k^* i-2\eta}{2}=\prod^{2N+4}_{k=1}\sinh\frac{u^*-z_k i-2\eta}{2},
\eea
which means that if $z_k$ is one zero root of $\Lambda^{s+}(u)$, then $-z_k$, $z_k^*$ and $-z_k^*$ must be the roots.
Due to the periodicity of $\Lambda^{s+}(u)$, we fix the real parts of $\{z_{j}\}$ in the interval $(-\pi, \pi]$.

We focus that $\{\theta_k\}$ are pure imaginary. The inhomogeneity parameters are imaginary to keep the transfer matrix $t^{s+}(u)$ hermitian for real $\eta$ case. From the numerical solutions of Eq.\eqref{sollf1}-\eqref{tipi} with finite system size $N$ and the singularity analysis in the thermodynamic limit,
we obtain the distribution of zero roots $\{z_j\}$ as follows.
(I) Real roots $\{z_j\}$.  (II) Bulk strings $\{z_j, z_j^*\}$, which means that two roots form the conjugate pairs with the imaginary parts around the points of $\pm2in\eta$ ($n\geq 2$).
We remark that the structures of bulk strings (II) are quite similar with the strings of Bethe roots for the periodic case.
(III) Boundary strings either at the origin or at $\text{Re}\{ z_j\}=\pi$, where the imaginary parts of conjugate pairs are determined by the boundary parameters.
The boundary strings are tightly related with the bound states induced by the boundary magnetic fields.
(IV) Additional roots, which means that a conjugate pair with imaginary parts neither around $\pm2in\eta$ ($n\geq 2$) nor having a simple relation with boundary parameters.
The additional roots are closely related with the Bethe-ansatz-like equations at the special points \eqref{t0}-\eqref{tipi}.

The above four kinds of zero roots are valid for the whole energy spectrum. From now, we focus on the ground state.
Only the configurations (I), (III) and (IV) of zero roots appear at the ground state.
The further analysis gives that the boundary strings take the form of
\bea
z^b_l=\pi\pm(2\eta-\alpha^{y}_{x})i,\,\,{\rm and}\,\,\alpha^{y}_{x}\in\{\alpha_{1},\alpha_{2},\alpha'_{1},\alpha'_{2}\}, \label{aa2}
\eea
where parameters $\alpha_{1},\alpha_{2},\alpha'_{1},\alpha'_{2}$ are defined in Eq.\eqref{alpha}.
We should note that the fixed boundary magnetic fields can give the positive or negative values of
$\text{Re}(\alpha^{y}_{x})$ due to the property of hyperbolic cosine function. Please see the parametrization \eqref{alpha}.
However, from Eq.\eqref{aa2} we know that if the signs of $\text{Re}(\alpha^{y}_{x})$ are different,
the lengthes of corresponding boundary strings are different.
Thus we need some selection rules to give the correct boundary strings. If $h_1^z h_N^z>0$, the
selection rule is that all the $\text{Re}(\alpha^{y}_{x})$ take the positive value. While if $h_1^z h_N^z<0$,
the selection rule is that the smaller $\text{Re}(\alpha^{y}_{x})$ in the set $\alpha^{y}_{x} \in \{\alpha_1,\alpha'_1\}$ takes the negative value and the others remain positive.

Another thing we should remark is that if $\text{Re}(\alpha^y_x)<2\eta$, there exist the boundary strings determined by the boundary parameter $\alpha^y_x$.
If $\text{Re}(\alpha^y_x)>2\eta$, the corresponding boundary string vanishes.
Because $\alpha^{y}_{x}\in\{\alpha_{1},\alpha_{2},\alpha'_{1},\alpha'_{2}\}$, the number of boundary strings varies from zero to four.

\begin{figure}[ht]
\centering
\includegraphics[width=8cm,height=6cm]{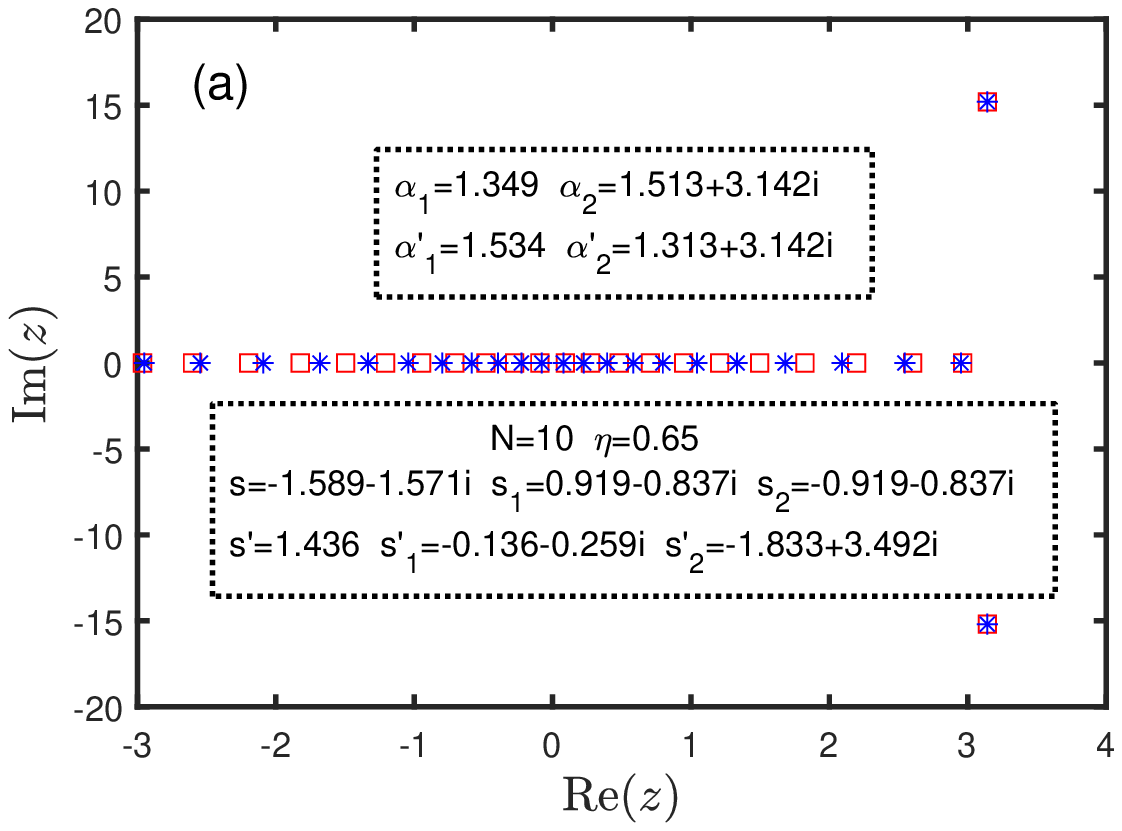}
\includegraphics[width=8cm,height=6cm]{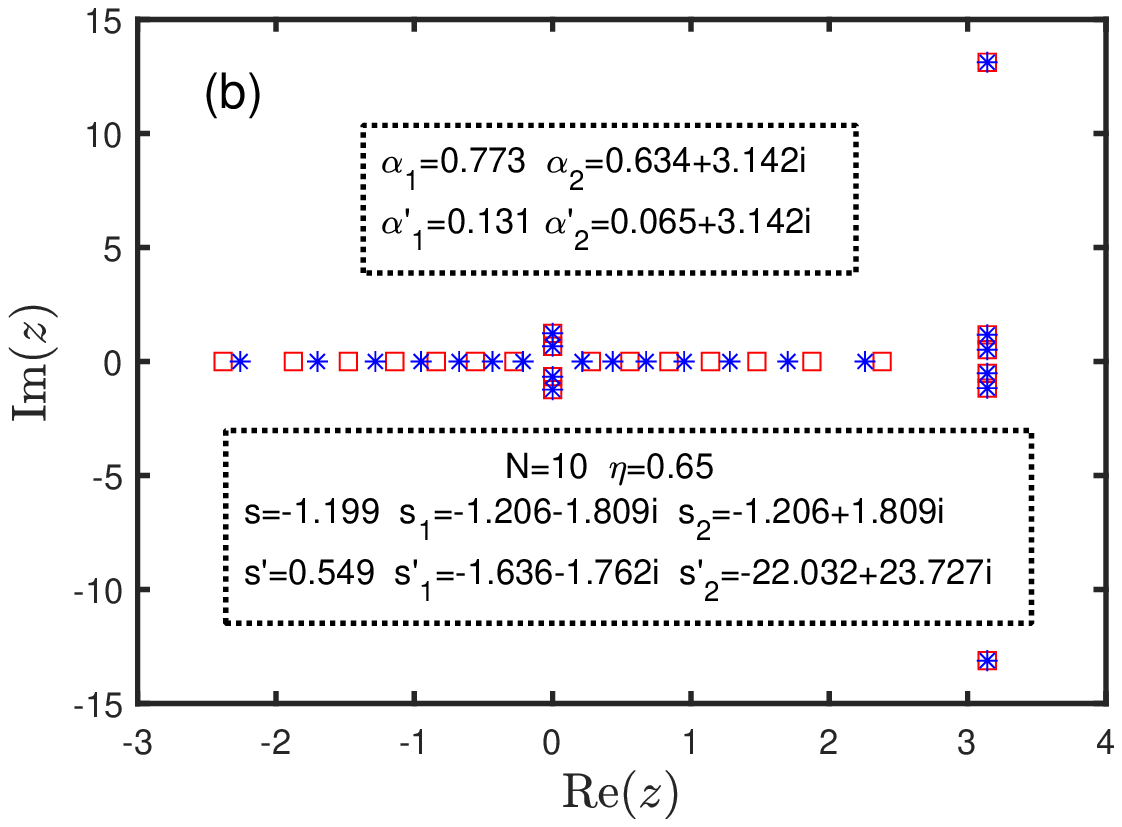}
\caption{The distributions of zero roots with certain model parameters at the ground state. (a) The pattern includes the real roots and two additional roots at the boundary. (b) The pattern includes the real roots, two boundary strings at the origin, two boundary strings and two additional roots at the boundary.
The red squares specify the roots of Bethe-ansatz-like equations \eqref{sollf1}-\eqref{tipi} with the inhomogeneity parameters $\{\bar{\theta}_k=-i\theta_k=0.13k\}$ and
the blue asterisks indicate the roots in the homogeneous limit obtained by the exact diagonalization of Hamiltonian.
We see that the inhomogeneity parameters do not affect the imaginary parts of roots but the positions in the real axis.}\label{Eg-theta}
\end{figure}

The numerical solutions of the Bethe-ansatz-like equations \eqref{sollf1}-\eqref{tipi} and exact numerical diagonalization results with certain model parameters at the ground state are shown in Fig.\ref{Eg-theta}. Fig.\ref{Eg-theta}(a) shows the case that there is no the boundary string. The pattern of zero roots includes the real roots and
two additional roots at the boundary.
The pattern of Fig.\ref{Eg-theta}(b) includes the real roots, four boundary strings (two of them at the origin and two at the boundary) and two additional roots at the boundary.
We note that the imaginary inhomogeneity parameters almost do not affect the imaginary parts of zero roots but the positions in the real axis.

\subsection{Surface energy of $XXZ$ spin-1/2 chain}

We first consider the case that there is no the boundary string, where all $\text{Re}(\alpha_x^y)>2\eta$.
The zero roots at the ground state include the real roots $\{\bar{z}_k\}$ and the additional roots
$\{\pi \pm z_a i\}$, where $z_a$ is real.
In the thermodynamic limit, where the system size $N$ tends to infinity, the distribution of zero roots can be characterized by a density per site $\rho(\bar{z})$.
Furthermore, we assume that the inhomogeneity parameters also has a continuum density per site $\sigma(\bar{\theta}_j)\sim 1/N(\bar{\theta}_j-\bar{\theta}_{j-1})$, where $\bar{\theta}_j=-i\theta_j$.
Thus the density of zero roots $\rho(\bar{z})$ can be derived with the help of an auxiliary function $\sigma(\bar{\theta})$ which is a given density of the inhomogeneity.

Substituting the pattern of zero roots at the ground state into the Bethe-ansatz-like equations \eqref{sollf1}, taking the logarithm,
making the difference of the resulted equations for $\theta_j$ and $\theta_{j-1}$, in the thermodynamic limit, we obtain
\bea
&& N\int_{-\pi}^{\pi}b_2(u-\bar{z})\rho(\bar{z})d\bar{z}  +b_{\frac{z_a}\eta+2}(u-\pi)+b_{\frac{z_a}\eta-2}(u-\pi) \no \\ [4pt]
&& = N\int_{-\pi}^{\pi}[b_4(u-\bar{\theta})+b_4(u+\bar{\theta})] \sigma(\bar{\theta})d\bar{\theta}+ b_4(u-\pi)+b_4(u) \no \\ [4pt]
&& -b_2(u)-b_2(u-\pi) +b_{\frac{\alpha_{1}}{\eta}}(u-\pi)+b_{\frac{\alpha_{1}'}{\eta}}(u-\pi) +b_{\frac{\bar{\alpha}_{2}}{\eta}}(u)+b_{\frac{\bar{\alpha}_{2}'}{\eta}}(u),\label{ai}
\eea
where the functions $a_n(u)$ and $b_n(u)$ are defined as
\bea
&&a_n(u)=(\ln\sin\frac{u+n\eta i}2-\ln\sin\frac{u-n\eta i}2)'=\frac12(\cot\frac{u+n\eta i}2-\cot\frac{u-n\eta i}2),
\no \\ [4pt]
&&b_n(u)=(\ln\sin\frac{u+n\eta i}2+\ln\sin\frac{u-n\eta i}2)'=\frac12(\cot\frac{u+n\eta i}2+\cot\frac{u-n\eta i}2).
\eea
Eq.\eqref{ai} is a standard convolution integral equation and can be solved by the Fourier transformation.
The Fourier transformations of Eq.\eqref{ai} is
\bea
&& N \tilde{b}_2(k)\tilde{\rho}(k) +\tilde{b}_{\frac{z_a}\eta+2}(k)e^{-ik\pi}+\tilde{b}_{\frac{z_a}\eta-2}(k)e^{-ik\pi}= 2N\tilde{b}_4(k)\tilde{\sigma}(k)+ \tilde{b}_4(k)e^{ik\pi}+\tilde{b}_4(k) \no \\ [4pt]
&&-\tilde{b}_2(k)-\tilde{b}_2(k)e^{-ik\pi} +\tilde{b}_{\frac{\alpha_{1}}{\eta}}(k)e^{-ik\pi} +\tilde{b}_{\frac{\alpha_{1}'}{\eta}}(k)e^{-ik\pi} +\tilde{b}_{\frac{\bar{\alpha}_{2}}{\eta}}(k) +\tilde{b}_{\frac{\bar{\alpha}_{2}'}{\eta}}(k),\label{rhoeq}
\eea
where $\tilde{a}_n(k)$ and $\tilde{b}_n(k)$ are calculated as
\bea
&&\tilde{a}_n(k)=\int_{-\pi}^{\pi}a_n(u)e^{-iku} du =-2\pi i \, \text{sign}(k)e^{-|n\eta k|},
\no \\ [4pt]
&&\tilde{b}_n(k)=\int_{-\pi}^{\pi}b_n(u)e^{-iku} du=-2\pi i \, \text{sign}(n\eta)e^{-|n\eta k|}.
\eea
In the homogeneous limit, we take $\sigma(\bar{\theta})=\delta(\bar{\theta})$. Thus $\tilde{\sigma}(k)=1$.
From Eq.\eqref{rhoeq}, we obtain the density of zero roots as
\bea
\tilde{\rho}(k) &=& 2e^{-2\eta|k|}+ \frac{1}{N}[ e^{-2\eta|k|}e^{ik\pi}+e^{-2\eta|k|}-1-e^{-ik\pi} +e^{-(\alpha_{1}-2\eta)|k|}e^{-ik\pi}+e^{-(\alpha_{1}'-2\eta)|k|}e^{-ik\pi} \no \\ [4pt]
&& +e^{-(\bar{\alpha}_{2}-2\eta)|k|} +e^{-(\bar{\alpha}_{2}'-2\eta)|k|} -e^{-z_a|k|}e^{-ik\pi} -e^{-(z_a-4\eta)|k|}e^{-ik\pi}].\label{rhoeq1}
\eea
The inverse Fourier transformation gives
\bea
\rho(u) &=& \frac{i}{\pi} a_2(u)+ \frac{i}{2\pi N}[ a_2(u-\pi)+a_2(u)+a_{\frac{\alpha_{1}}{\eta}-2}(u-\pi) +a_{\frac{\alpha_{1}'}{\eta}-2}(u-\pi)+a_{\frac{\bar{\alpha}_{2}}{\eta}-2}(u) \no \\
&&+a_{\frac{\bar{\alpha}_{2}'}{\eta}-2}(u) -a_{\frac{z_a}\eta}(u-\pi)-a_{\frac{z_a}\eta-4}(u-\pi)] -\frac{1}{N}[\delta(u)+\delta(u-\pi)].\label{rhoeq2}
\eea
Thus the ground state energy of the Hamiltonian \eqref{Hs1} is
\bea
E^{s+}_g &=&-\frac N4\int_{-\pi}^{\pi}\coth\frac{2\eta+z i}2 \rho(z)dz +\frac{\tanh(2\eta)}4(1+\coth s') \no \\[4pt] && -\frac14[\tanh\frac{2\eta+z_a}2+\tanh\frac{2\eta-z_a}2] \no \\[4pt] &=&
-\frac N2 \coth(2\eta)-\frac14[\coth(2\eta)-\coth\eta-\tanh\eta +\tanh\frac{\alpha_1}2+\tanh\frac{\alpha'_1}2 \no \\[4pt] && +\coth\frac{\bar{\alpha}_2}2 +\coth\frac{\bar{\alpha}'_2}2-\tanh(2\eta)\coth s'+2\tanh\frac{2\eta-z_a}2].\label{Esg}
\eea

Now, we prove that the $z_a$ tends to infinity in the thermodynamic limit.
Taking the logarithm of Eq.\eqref{t0} and considering the homogeneous limit $\{\theta_j=0\}$, we have
\bea
&&\log\Lambda_0+N\log(-1)+\sum^{2N+4}_{k=1}\log\sin\frac{z_k-2\eta i}{2}=\log2+\log\cosh(2\eta) \no \\ && +\log\sinh s+\log\sinh s'+2N\log\sinh(2\eta).\label{t01}
\eea
In the thermodynamic limit, Eq.\eqref{t01} reads
\bea
&&\log\Lambda_0+N\log(-1)+N\int^{\pi}_{-\pi}\log\sin\frac{\bar{z}-2\eta i}{2}\rho(\bar{z})d\bar{z}+\log\cosh\frac{2\eta-z_a}2+\log\cosh\frac{2\eta+z_a}2 \no \\[4pt]
 && =\log2+\log\cosh(2\eta)+\log\sinh s+\log\sinh s'+2N\log\sinh(2\eta).\label{t02}
\eea
Substituting the density \eqref{rhoeq2} into \eqref{t02} and omitting the $O(N^{-1})$ terms, we have
\bea
z_a&=&2N\log\sinh(2 \eta)-\frac{iN}{\pi}\int_{-\pi}^{\pi}\log\sin\frac{\bar{z}-2\eta i}{2} a_2(\bar{z})d\bar{z}+C \no \\ &=&2\eta N+C',\label{za1}
\eea
where $C'$ is a constant irrelevant with $N$. Thus $z_a$ tends to infinity if $N\rightarrow \infty$.
The same result can also be obtained by using the Bethe-ansatz-like equations \eqref{tipi} with the similar method.

\begin{figure}[ht]
\centering
\includegraphics[width=8cm,height=5cm]{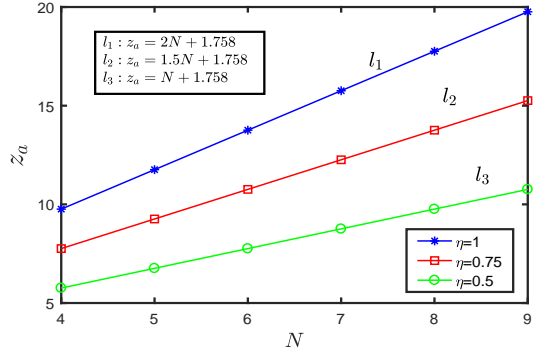}
\caption{The finite size scaling behaviors of $z_a$. The blue asterisks, red squares and green circles are the numerical solutions of $z_a$ with $\eta=0.5$, $0.75$ and $1$, respectively. The data can be fitted as $a N+b$. It is clear that $a=2\eta$, $b=1.758$, and $z_a \rightarrow \infty$ if $N \rightarrow \infty$.
 Here the model parameters are $h_1^+=0.23+0.36i$, $h_1^-=0.23-0.36i$, $h_1^z=1.2$, $h_N^+=0.82+0.93i$, $h_N^-=0.82-0.93i$ and $h_N^z=3.23$. }\label{zaf}
\end{figure}

Next, we check the conclusion \eqref{za1} numerically. By using the exact numerical
diagonalization method, we obtain the values of $z_a$ with certain model parameters.
In the homogeneous limit, the values of $z_a$ for different system sizes are shown in Fig.\ref{zaf}.
The data can be fitted as $a N+b$. From the fitting, we see that $a=2\eta$ and $b=1.758$ which is irrelevant with $N$.

Now, we are ready to calculate the surface energy of the Hamiltonian \eqref{Hs1}.
The surface energy is defined by
\bea
E^{s+}_s = E^{s+}_g-E^{sp}_{g},\label{1Ess}
\eea
where $E^{s+}_g$ is the ground state energy given by \eqref{Esg} and
\bea
E^{sp}_{g}=-\frac{N}{2}\coth 2\eta,
\eea
is the ground state energy of the system with periodic boundary condition
\bea H^{sp}=-\sum_{j=1}^{N}
\frac{1}{4\sinh2\eta}[\sigma^x_j\sigma^x_{j+1}+\sigma^y_j\sigma^y_{j+1}+\cosh(2\eta)
\sigma^z_j\sigma^z_{j+1}]-\frac{N}{4}\coth 2\eta.
\eea
Considering $z_a\rightarrow\infty$ in the thermodynamic limit, we obtain the surface energy as
\bea
E^{s+}_s &=& \frac12-\frac14[\coth(2\eta)-\coth\eta-\tanh\eta +\tanh\frac{\alpha_1}2+\tanh\frac{\alpha'_1}2 \no \\ && +\coth\frac{\bar{\alpha}_2}2 +\coth\frac{\bar{\alpha}'_2}2-\tanh(2\eta)\coth s'].\label{Ess}
\eea

Next, we consider the case that there exists one boundary string. For example, we choose the boundary parameters in the regime of $\alpha_1<2\eta$, $\alpha_1^\prime>2\eta$, $\text{Re}(\alpha_2)>2\eta$ and $\text{Re}(\alpha_2^\prime)>2\eta$.
In this case, only the boundary string $\pi\pm(2\eta-\alpha_1)i$ exists.
This boundary string would affect the distribution of zero roots and make the density $\rho(z)$ has a deviation.
After some calculations, we obtain the deviation $\delta\tilde{\rho}(k)$ as
\bea
\delta\tilde{\rho}(k)=-\frac1N[e^{-(2\eta-\alpha_1)|k|}+e^{(2\eta-\alpha_1)|k|}]e^{-ik\pi}.
\eea
The contribution of boundary string to the ground state energy is
\bea
\delta E^{s+}_g=-\frac N4\int_{-\pi}^{\pi}\coth\frac{2\eta+z i}2 \delta\rho(z)dz-\frac14 (\tanh\frac{\alpha_1}2+\tanh\frac{4\eta-\alpha_1}2)=0.
\eea
Thus the boundary string contributes nothing to the ground state energy.
The ground state energy \eqref{Esg} and surface energy \eqref{Ess} are correct for the all regimes of boundary parameters.

The ground state energy of the Hamiltonian $H^{s-}$ can be obtained similarly
\bea
E^{s-}_g=E^{s+}_g|_{\{\alpha_1,\alpha'_1,\alpha_2,\alpha'_2\}\rightarrow \{\gamma_1,\gamma'_1,\gamma_2,\gamma'_2\}},
\eea
where the boundary parameters are defined as
\bea
&&\hspace{-1.0truecm}\gamma=\frac{1}{2t_1t_2},\quad
\cosh\gamma_1=\frac{\gamma}{2}+\xi, \quad \cosh\gamma_2=\frac{\gamma}{2}-\xi, \quad \xi=\sqrt{1+ \frac{\gamma^2}{4}+\gamma\cosh(2t)},\no\\[8pt]
&&\hspace{-1.0truecm}\gamma'=\frac{1}{2t'_1t'_2},\,\, \cosh\gamma'_1=\frac{\gamma'}{2}+\xi', \,\, \cosh\gamma'_2=\frac{\gamma'}{2}-\xi', \quad \xi'=\sqrt{1+\frac{\gamma'^2}{4}+\gamma'\cosh(2t')}.\label{gamma}
\eea
The selection rules of $\gamma_x^y\in\{\gamma_1,\gamma'_1,\gamma_2,\gamma'_2\}$ are
that all the $\text{Re}(\gamma^{y}_{x})$ take positive values if $\coth s \coth s'<0$; or the smaller $\text{Re}(\gamma^{y}_{x})$ in the set $\gamma^{y}_{x} \in \{\gamma_1,\gamma'_1\}$ takes the negative value and the others remain positive if $\coth s \coth s' >0$. Similar to $\alpha_2$ and $\alpha_2'$, we denote $\gamma_2=\bar{\gamma}_2+\pi i$ and $\gamma'_2=\bar{\gamma}'_2+\pi i$, where $\bar{\gamma}_2$ and $\bar{\gamma}'_2$ are real.

\subsection{Surface energy of the anisotropic $D^{(1)}_2$ spin chain}

Now, we are ready to calculate the surface energy of the anisotropic $D^{(1)}_2$ spin chain.
The ground state energy of the system \eqref{apopd} with the twisted boundary magnetic fields is
\bea
E_g &=&-N\coth(2\eta)+1-\frac14[2\coth(2\eta)-2\coth\eta-2\tanh\eta +\tanh\frac{\alpha_1}2+\tanh\frac{\alpha'_1}2 \no \\ [4pt] && +\coth\frac{\bar{\alpha}_2}2 +\coth\frac{\bar{\alpha}'_2}2-\tanh(2\eta)\coth s'
+\tanh\frac{\gamma_1}2+\tanh\frac{\gamma'_1}2 \no \\ [4pt] && +\coth\frac{\bar{\gamma}_2}2 +\coth\frac{\bar{\gamma}'_2}2-\tanh(2\eta)\coth t'].
\eea
The Hamiltonian of anisotropic $D^{(1)}_2$ spin chain with periodic boundary condition is
\bea&&H^p=-\sum_{j=1}^{N}\frac{1}{4\sinh(2\eta)}\{\cosh
(2\eta)(\sigma^z_j\sigma^z_{j+1}+\tau^z_j\tau^z_{j+1})+2[
\sigma^z_j\sigma^z_{j+1}(\tau^+_j\tau^-_{j+1}+\tau^-_j\tau^+_{j+1})\no\\
&&\qquad +(\sigma^+_j\sigma^-_{j+1}
+\sigma^-_j\sigma^+_{j+1})\tau^z_j\tau^z_{j+1}]\}-
\frac N2\coth(2\eta).\label{p1o}\eea
The ground state energy of the Hamiltonian \eqref{p1o} is
\bea
E^p_{g}=-N\coth(2\eta).
\eea
Thus the surface energy of anisotropic $D^{(1)}_2$ spin chain reads
\bea
E_s &=& 1-\frac14[2\coth(2\eta)-2\coth\eta-2\tanh\eta +\tanh\frac{\alpha_1}2+\tanh\frac{\alpha'_1}2 \no \\ && +\coth\frac{\bar{\alpha}_2}2 +\coth\frac{\bar{\alpha}'_2}2-\tanh(2\eta)\coth s'
+\tanh\frac{\gamma_1}2+\tanh\frac{\gamma'_1}2 \no \\ && +\coth\frac{\bar{\gamma}_2}2 +\coth\frac{\bar{\gamma}'_2}2-\tanh(2\eta)\coth t'],\label{Esaniso}
\eea
where the parameters $\alpha,\alpha_1, etc.$ are given by \eqref{alpha} and \eqref{gamma}. The energy expression \eqref{Esaniso} is valid for $\eta>0$ and all the other fundamental parameters ($s, s_1, s_2, etc.$) which keep the Hamiltonian (4.15) hermitian (eg. $s=s^*$, $s_1=s_2^*$, $s'=s'^*$, $s_1'=e^{(s_r+i s_i)\eta}$, $s_2'=e^{(s_r+4-i s_i)\eta}$, $t=t^*$, $t_1=t_2^*$, $t'=t'^*$, $t_1'=e^{(t_r+i t_i)\eta}$, $t_2'=e^{(t_r+4-i t_i)\eta}$, where $s_r, s_i, t_r, t_i$ are real, and the superscript * means the complex conjugate).

\begin{figure}[ht]
\centering
\includegraphics[width=8cm,height=5cm]{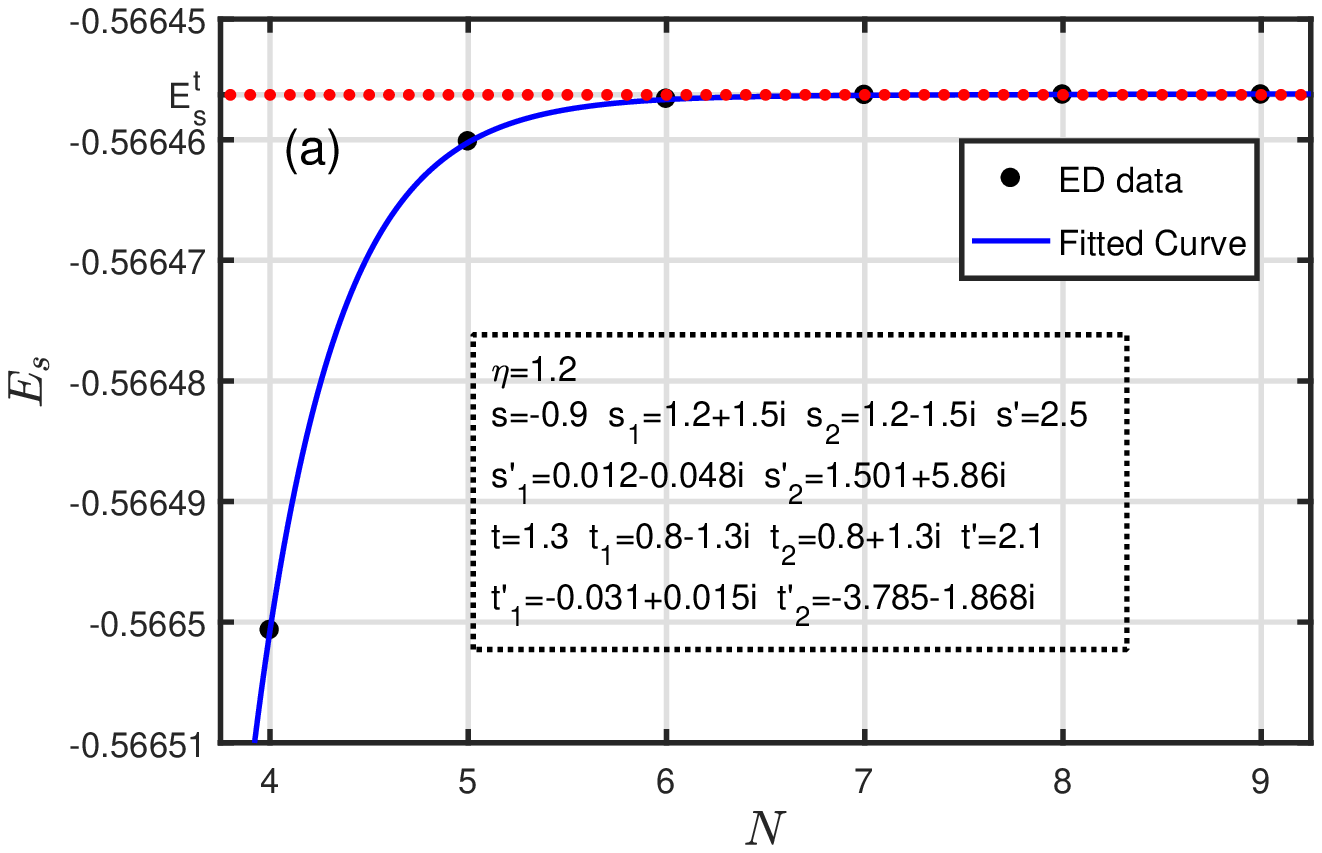}
\includegraphics[width=8cm,height=5cm]{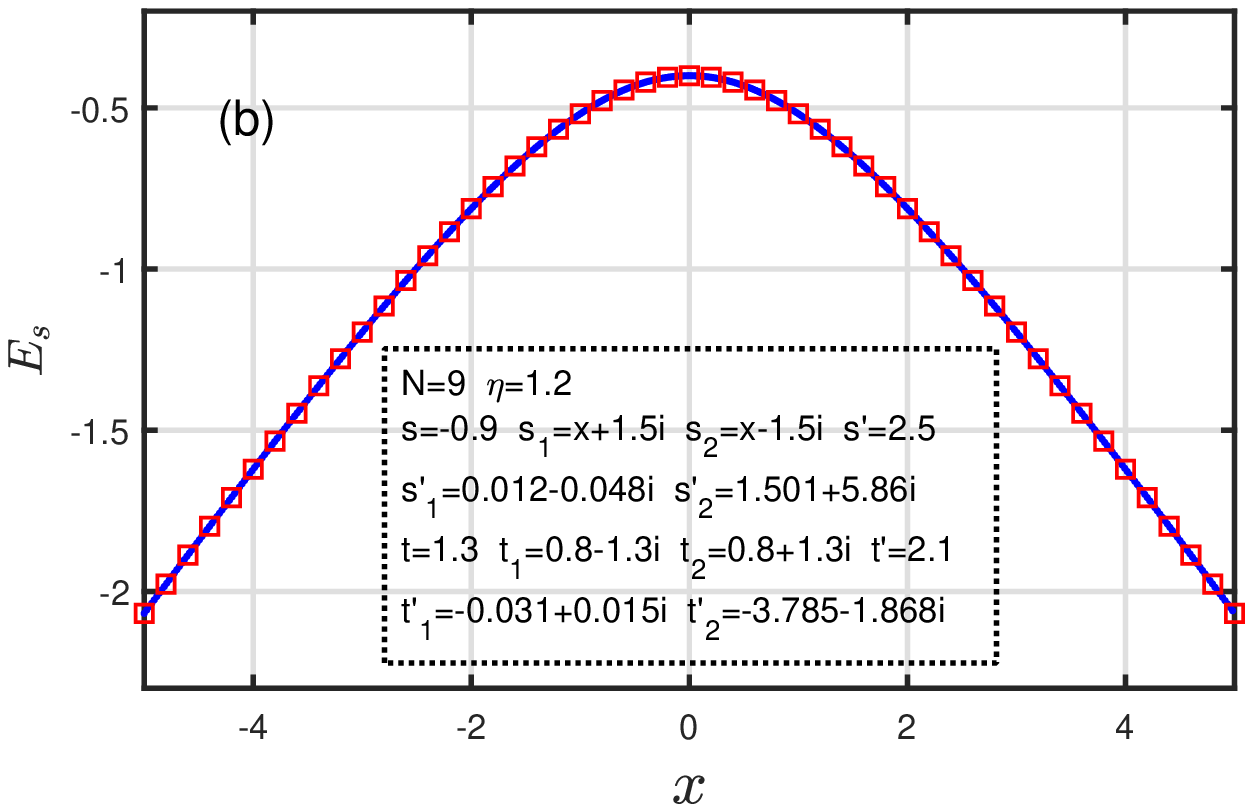}
\caption{(a) The finite size scaling behavior of the surface energy $E_s$ of the anisotropic $D^{(1)}_2$ spin chain. The black points are calculated from the exact diagonalization method with the finite system size $N$. The data can be fitted as $E_s= -0.778485e^{-2.44378N}-0.566456$. When $N$ tends to infinity, the finite size scaling analysis gives the value $-0.566456$, which is exactly the same as the surface energy calculated from the analytic expression \eqref{Esaniso}.
(b) The surface energy $E_s$ versus the boundary parameter $x=\text{Re}(s_1)$. The blue line indicates the analytic result and the red squares indicate the numerical results.
 They are consistent with each other very well. }\label{Es-diff}
\end{figure}

We check the conclusion \eqref{Esaniso} numerically. The surface energy can also be calculated by using the exact diagonalization method with the help of finite size scaling analysis.
Here, the exact diagonalization is performed with the anisotropic $\eta=1.2$ and the system size $N$ is set from 4 to 9.
The results are shown in Fig.\ref{Es-diff}. The exact diagonalization results with the fixed boundary parameters are shown in Fig.\ref{Es-diff}(a).
The date can be fitted as $E_s= -0.778485e^{-2.44378N}-0.566456$. When $N$ tends to infinity,
the finite size scaling analysis gives the value $-0.566456$, which is exactly the same as the surface energy calculated from the analytic expression \eqref{Esaniso}.
In Fig.\ref{Es-diff}(b), we show the surface energy versus one free boundary parameter. Again, the analytic results obtained from \eqref{Esaniso}
and the numerical ones calculated from exact diagonalization are consistent with each other very well.

\section{Results for the isotropic $D^{(1)}_2$ spin chain}
\setcounter{equation}{0}

\subsection{Exact solution}
The isotropic $D^{(1)}_2$ spin chain is quantified by the transfer matrix
\begin{equation}
\tilde t(u)= tr_0 \{\bar {\tilde K}_0(u) \tilde R_{01}(u-\theta_1)\cdots \tilde R_{0N}(u-\theta_N) \tilde K_0(u)
\tilde R_{N0}(u+\theta_N) \cdots \tilde R_{10}(u+\theta_1)\}. \label{ai8}
\end{equation}
Here the dual reflection matrix $\bar {\tilde K}(u)$ can be obtained from the reflection matrix $\tilde K(u)$ by the mapping
\bea
\bar {\tilde K}(u)= \tilde K(-u-1)|_{\{t, t_1, t_2, s, s_1, s_2\}\rightarrow \{t', t'_1, t'_2, s', s'_1, s'_2\}},  \label{ai9}
\eea
and $\tilde K(u)$ is the $4\times 4$ matrix with the elements
\bea
&& \tilde k_{11}(u)=(s-u)(t-u),\quad \tilde k_{12}(u)=-t_1u(s-u),\quad \tilde k_{13}(u)=s_1u(t-u),\no\\
&& \tilde k_{14}(u)=s_1t_1u^2,\quad \tilde k_{21}(u)=-t_2u(s-u),\quad \tilde k_{22}(u)=(s-u)(t+u),\no\\
&& \tilde k_{23}(u)=-s_1t_2u^2,\quad \tilde k_{24}(u)=-s_1u(t+u),\quad \tilde k_{31}(u)=s_2u(t-u),\no\\
&& \tilde k_{32}(u)=-s_2t_1u^2,\quad  \tilde k_{33}(u)=(s+u)(t-u),\quad\tilde  k_{34}(u)=t_1u(s+u),\no\\
&& \tilde k_{41}(u)=s_2t_2u^2,\quad \tilde  k_{42}(u)=-s_2u(t-u),\quad\tilde  k_{43}(u)=t_2u(s+u),\no\\
&&\tilde  k_{44}(u)=(s+u)(t+u). \label{ai10} \eea
The associated $R$-matrix reads
\bea
\tilde R_{12}(u)=\left(\begin{array}{cccc|cccc|cccc|cccc}
    \tilde a&&& &&&& &&&& &&&& \\
    &\tilde b&& &\tilde g&&& &&&& &&&& \\
    &&\tilde b& &&&& &\tilde g&&& &&&& \\
    &&&\tilde e &&&\tilde d& &&\tilde d&& &\tilde c&&& \\
   \hline &\tilde g&& &\tilde b&&& &&&& &&&& \\
    &&& &&\tilde a&& &&&& &&&& \\
    &&&\tilde d &&&\tilde e& &&\tilde c&& &\tilde d&&& \\
    &&& &&&&\tilde b &&&& &&\tilde g&& \\
   \hline &&\tilde g& &&&& &\tilde b&&& &&&& \\
    &&&\tilde d &&&\tilde c& &&\tilde e&& &\tilde d&&& \\
    &&& &&&& &&&\tilde a & &&&& \\
    &&& &&&&  &&&&\tilde b &&&g& \\
   \hline &&&\tilde c &&&\tilde d& &&\tilde d&& &\tilde e&&& \\
    &&& &&&&\tilde g &&&& &&\tilde b&& \\
    &&& &&&& &&&&\tilde g &&&\tilde b& \\
    &&& &&&& &&&& &&&&\tilde a \\
           \end{array}\right),\label{RD-msdatrix}
\eea where the matrix elements are \bea
\tilde a=(u+1)(u+1), \;\;
\tilde b=u(u+1), \;\;\tilde  c=1,\;\;\tilde  d=-u,\;\; \tilde e=u^2,\;\; \tilde g=u+1. \no \eea

The reflection matrices \eqref{ai9}-\eqref{ai10} can be factorized as
\bea
\bar {\tilde K}(u)=S [\bar{\tilde K}^{s+}(u) \otimes \bar{\tilde K}^{s-}(u)]S^{-1}, \quad
\tilde K(u)=S[ \tilde K^{s+}(u) \otimes \tilde K^{s-}(u)]S^{-1},
\eea
where the matrix $S$ is given by Eq.\eqref{ai7} and
\bea && \bar {\tilde K}^{s+}(u)= \tilde K^{s+}(-u-1)|_{\{s,s_1,s_2\}\rightarrow\{s',s'_1,s'_2\}}, \quad
\tilde K^{s+}(u)=\left(\begin{array}{cc}
    s-u&s_1 u \\
   s_2 u &s+u
   \end{array}\right), \label{aaao}\\
&& \bar {\tilde K}^{s-}(u)= \tilde
K^{s+}(-u-1)|_{\{t,t_1,t_2\}\rightarrow\{ t',t'_1,t'_2\}}, \quad
\tilde K^{s-}(u)=\tilde K^{s-}(u)|_{\{s,s_1,s_2\}\rightarrow\{
t,t_1,t_2\}}. \eea The $R$-matrix \eqref{RD-msdatrix} can be
decomposed into \bea \tilde R_{12}(u)=[S\otimes S] [\tilde
R^{s}_{1'2'}(u)\otimes \tilde R^{s}_{1''2''}(u)] [S\otimes
S]^{-1}, \eea where $\tilde R^{s}_{1'2'}(u)$ is the $R$-matrix of
the isotropic XXX spin chain \bea \tilde
R^{s}_{1'2'}(u)=\left(\begin{array}{cccc}
    u+1&0&0&0 \\
   0 &u&1&0 \\
    0&1&u&0 \\
   0 &0&0&u+1
   \end{array}\right).\label{Rxxx}
\eea

Based on the above decompositions, we obtain the eigenvalue of
transfer matrix \eqref{ai8} as \bea \tilde \Lambda(u)=\tilde
\Lambda^{s+}(u)\tilde \Lambda^{s-}(u),\label{1BA012}\eea where
\bea &&\tilde \Lambda^{s+}(u)=\frac{u+1}
{u-\frac12}(s+\sqrt{1+s_1s_2}u)(s'-\sqrt{1+s'_1s'_2}u)\tilde a(u)\frac{\tilde Q(u-1)}{\tilde Q(u)}\no\\[4pt]
&&\quad\quad\quad+\frac{u}
{u-\frac12}[s-\sqrt{1+s_1s_2}(u+1)][s'+\sqrt{1+s'_1s'_2}(u+1)]\tilde
d(u) \frac{\tilde Q(u+1)}{\tilde Q(u)}\no\\[4pt]
&&\quad\quad\quad+\tilde x_+u(u+1)\frac{\tilde a(u)\tilde d(u)}
{\tilde Q(u)},\no\\[4pt]
\hspace{-0.8truecm}&&\hspace{-0.8truecm}\tilde \Lambda^{s-}(u)=\tilde \Lambda^{s+}(u)|_{\{s, s_1, s_2, s', s'_1, s'_2\}\rightarrow \{t, t_1, t_2, t', t'_1, t'_2\}}, \label{tser}\eea
and
\bea
&&\tilde Q(u)=\prod_{l=1}^{N}(u-\mu_l)(u+\mu_l+1), \no \\
&&\tilde a(u)=\prod_{j=1}^N(u-\theta_j+1)(u+\theta_j+1)=\tilde d(u+1),  \no \\
&&\tilde x_+=2\sqrt{(1+s_1s_2)(1+s'_1s'_2)}-(2+s_1s'_2+s_2s'_1).\eea
The Bethe roots $\{u_l\}$ in Eq.\eqref{tser} should satisfy the Bethe ansatz equations
\bea
&&\frac{\mu_l+1} {\mu_l-\frac12}\frac{(s+\sqrt{1+s_1s_2}\mu_l)(s'-\sqrt{1+s'_1s'_2}\mu_l)}
{\tilde d(u_l)}{\tilde Q(\mu_l-1)}\no\\[4pt]
&&+\frac{\mu_l} {\mu_l-\frac12}\frac{[s-\sqrt{1+s_1s_2}(\mu_l+1)[[s'+\sqrt{1+s'_1s'_2}(\mu_l+1)]}
{\tilde a(u_l)}{\tilde Q(\mu_l+1)}\no\\[4pt]
&&=-\tilde x_+\mu_l(\mu_l+1), \quad l=1,\cdots, N. \label{BA012}\eea

The Hamiltonian of isotropic $D^{(1)}_2$ spin chain generated by the transfer matrix \eqref{ai8} is
\bea&&\tilde{H}=\sum_{j=1}^{N-1}[\frac12(\sigma^z_j\sigma^z_{j+1}+\tau^z_j\tau^z_{j+1})+
\sigma^z_j\sigma^z_{j+1}(\tau^+_j\tau^-_{j+1}+\tau^-_j\tau^+_{j+1})+(\sigma^+_j\sigma^-_{j+1}
+\sigma^-_j\sigma^+_{j+1})\tau^z_j\tau^z_{j+1}]
\no\\[4pt]
&&\qquad -\frac{1}{2st}(
s\tau^z_{N}+t\sigma^z_{N}+st_1\sigma^z_{N}\tau^+_{N}-s_1t\sigma^+_{N}\tau^z_{N}+st_2\sigma^z_{N}\tau^-_{N}-s_2t\sigma^-_{N}\tau^z_{N})\no\\[4pt]
&&\qquad+\frac{1}{2s't'}(
s'\tau^z_{1}+t'\sigma^z_{1}+s't'_1\sigma^z_{1}\tau^+_{1}-s'_1t'\sigma^+_{1}\tau^z_{1}+s't'_2\sigma^z_{1}\tau^-_{1}+s'_2t'\sigma^-_{1}\tau^z_{1})+N.
\label{apop} \eea
We find that the
Hamiltonian \eqref{apop} is the direct summation of two XXX
spin-1/2 chains with non-diagonal boundary magnetic fields up to the
similar transformation ${\cal S}$ \bea \tilde{H}={\cal S} [\tilde{H}^{s+}\oplus
\tilde{H}^{s-}]{\cal S}^{-1}, \label{apop1}\eea where \bea
&&\tilde{H}^{s+}=\sum_{j=1}^{N-1}\frac12(\sigma^x_j\sigma^x_{j+1}+\sigma^y_j\sigma^y_{j+1}+
\sigma^z_j\sigma^z_{j+1})+ \frac{1}{4s}(s_1+s_2)\sigma^x_N
+\frac{i}{4s}(s_1-s_2) \sigma^y_{N} -\frac{1}{2s} \sigma^z_N\no\\
&&\quad\qquad -\frac{1}{4s'}(s'_1+s'_2)\sigma^x_1 -\frac{i}{4s'}(s'_1-s'_2) \sigma^y_{1} +\frac{1}{2s'} \sigma^z_1 +\frac{N}{2}, \label{H1rational}\\[6pt]
&&\tilde{H}^{s-}=\tilde{H}^{s+}|_{
\{\sigma_j^\alpha, s, s_1, s_2, s', s'_1, s'_2\}\rightarrow \{\tau_j^\alpha, t, t_1, t_2, t',
t'_1, t'_2\}}, \eea with ${\cal{S}}=\otimes^N S$ and
$S=\frac{1}{2}(1-\sigma^z+\tau^z+\sigma^z\tau^z)$. The conclusion
\eqref{apop1} is consistent with the fact that the corresponding
$R$-matrix and reflection matrices in the transfer matrix $\tilde
t(u)$ of the system can be factorized. The eigenvalue of the
Hamiltonian \eqref{apop} is \bea \tilde{E}^{s+}=\frac{\partial \ln \tilde
\Lambda(u)}{2\partial u}|_{u=0,\{\theta_j\}=0}, \eea where $\tilde \Lambda(u)$
is given by Eq.\eqref{1BA012}. In the derivation, the identity $tr_0{\bar{\tilde{K}}_0}(0)'=0$ is used.

\subsection{Symmetry}

In the Hamiltonian \eqref{H1rational}, there are six free boundary parameters. However, due to the fact that the interactions in the bulk are isotropic
and have the $su(2)$ symmetry, the free boundary parameters can be reduced. The method is as follows.
First, we choose the direction of magnetic field at the left side as the $z$-direction, which can be achieved by acting the transformation matrix $\otimes_{j=1}^{N}\tilde{C}_j^1$ on the Hamiltonian \eqref{H1rational}.
Then, we take a rotation of the resulted Hamiltonian in the $xy$ plane to make the direction of magnetic field at the right side lying in the $xz$ plane.
This can be achieved by using the transition matrix $\otimes_{j=1}^{N} \tilde{C}^2_j$.
We note that the interactions in the bulk do not change after the transformations.

The final result is
\bea
\bar{H}^{s+}&=& \otimes_{j=1}^{N} [\tilde C^2_j]^{-1} [\tilde C^1_j]^{-1}\tilde{H}^{s+} \tilde C^1_j \tilde C^2_j \no \\ &=&\frac12\bigg[\sum_{j=1}^{N-1}(\sigma^x_j\sigma^x_{j+1}+\sigma^y_j\sigma^y_{j+1}+
\sigma^z_j\sigma^z_{j+1})+\frac1p\sigma^z_1 +\frac1q(\xi\sigma^x_N+\sigma^z_N)\bigg]+\frac{N}{2}.\label{H1_iso}
\eea
Here the transformation matrices $\tilde C^1_j$ and $\tilde C^2_j$ are
\bea \tilde C^1_j=\left(\begin{array}{cc}
    C^1_{11}&C^1_{12}\\
    C^1_{21}&C^1_{22}
   \end{array}\right)_{[j]},\quad
\tilde C^2_j=\left(\begin{array}{cc}
    C^2_{11}&0\\
    0&1
   \end{array}\right)_{[j]},
\eea
and the matrix elements are
\bea
C^1_{11}&\!\!\!=&\!\!\!\frac{s_1' \left(\bar{s}_1-1\right)}{\left| s_1'\right|  \sqrt{2\bar{s}_1(\bar{s}_1-1)}}, \qquad
C^1_{12}=-\frac{s_1' \sqrt{\bar{s}_1+1}}{\sqrt{2 \bar{s}_1} \left| s_1'\right| }, \qquad \bar{s}_1=\sqrt{\left| s_1'\right| ^2+1}, \no \\ [4pt]
C^1_{21}&\!\!\!=&\!\!\!\frac{\left| s_1'\right| }{\sqrt{2 \bar{s}_1(\bar{s}_1-1)}}, \qquad
C^1_{22}=\frac{\left| s_1'\right| }{\sqrt{2(\left| s_1'\right| ^2+\bar{s}_1+1)}},\no \\ [4pt]
C^2_{11}&\!\!\!=&\!\!\!s_1' s_1\sqrt{\bar{s}_1-\bar{s}_1^2}\left[s_1 \left(s_1'^*\right)^2 +2 \left(\bar{s}_1+1\right) s_1'^* -s_1^*\left(\bar{s}_1+1\right)^2 \right] \left[s_1'^*\left(\bar{s}_1^2+\bar{s}_1\right)\right]^{-1}
\no \\ [4pt] &\!\!\!\times&\!\!\!
\left\{s_1'^2 s_1^2 \left[s_1^* \left(s_1'^2 s_1^*+4 s_1'-4 s_1\right)+s_1^2 \left(s_1'^*\right)^2-2 s_1'^* \left(s_1' s_1 s_1^*+2 s_1'-2 s_1\right)\right]\right\}^{-\frac12}.
\eea
The boundary parameters $p$, $q$ and $\xi$ in the Hamiltonian \eqref{H1_iso} reads
\bea
p&\!\!\!=&\!\!\!- s'(\left| s_1'\right| ^2+1)^{-\frac12}, \qquad q=2 s (\left| s_1'\right| ^2+1)^{\frac12} (s_1 s_1'^*+s_1' s_1^*+2)^{-1}, \no \\ [4pt]
\xi&\!\!\!=&\!\!\!\left\{-s_1'^2 s_1^2 \left[s_1^* \left(s_1'^2 s_1^*+4 s_1'-4 s_1\right)+s_1^2 \left(s_1'^*\right)^2-2 s_1'^* \left(s_1' s_1 s_1^*+2 s_1'-2 s_1\right)\right]\right\}^{\frac12}\no \\ [4pt]
&& \times \left[s_1' s_1 \left(s_1 s_1'^*+s_1' s_1^*+2\right)\right]^{-1}.\label{pqxi}
\eea
We see that only three free boundary parameters are left. The hermitian of Hamiltonian \eqref{H1_iso} requires that all the $p$, $q$ and $\xi$ are real.
The Hamiltonians $\bar{H}^{s+}$ and $\tilde{H}^{s+}$ have the same eigen-energies. Thus in the following, we derive the eigenvalues of the Hamiltonians $\bar{H}^{s+}$
instead of $\tilde{H}^{s+}$.

\subsection{Surface energy for the isotropic $D^{(1)}_2$ Hamiltonian}

The Hamiltonian \eqref{H1_iso} is generated by the
transfer matrix $\bar t(u)$ as
\begin{equation}\label{Ham-def}
  \bar{H}^{s+} = \left. \frac{\partial \ln \bar{t} (u)}{2\partial u} \right|_{u = 0,
  \{\theta_j = 0\}},
\end{equation}
where $\bar t(u)$ is defined by
\bea\label{transfer-XXX-open}
\bar{t}(u)=tr_0\{K^+_0(u)\tilde{R}^s_{0,N}(u-\theta_N) \cdots \tilde{R}^s_{0,1}(u-\theta_1)K^-_0(u) \tilde{R}^s_{0,1}(u+\theta_1) \cdots \tilde{R}^s_{0,N}(u+\theta_N)\},
\eea
and the reflection matrices are
\begin{eqnarray}
  K^- (u) & = & \left( \begin{array}{ll}
    p + u & 0\\
    0 & p - u
  \end{array} \right), \\
  K^+ (u) & = & \left( \begin{array}{ll}
    q + u + 1 & \xi (u + 1)\\
    \xi (u + 1) & q - u - 1
  \end{array} \right).
\end{eqnarray}
From the definition \eqref{transfer-XXX-open}, we know that $\bar{t} (u)$ and its eigenvalue $\bar{\Lambda} (u)$
are the polynomials of $u$ with the degree $2 N + 2$.
Thus we parameterize the eigenvalue $\bar \Lambda (u)$ by its roots $\{z_j \}$ as
\begin{equation}\label{lambda}
  \bar{\Lambda} (u) = 2 \prod_{j = 1}^{2N + 2} \left( u - z_j i + \frac{1}{2} \right) .
\end{equation}
Acting Eq.\eqref{Ham-def} on a common eigenstate of $\bar{H}^{s+}$ and $\bar t(u)$,
we obtain the eigenvalue $\bar{E}^{s+}$ of the Hamiltonian $\bar{H}^{s+}$
\bea
\bar{E}^{s+}  =  \sum_{j = 1}^{2N + 2} \frac{1}{1 - 2 \hat{z}_j i},\label{XXXE}
\eea
where $\{\hat{z}_j\}$ are the homogeneous limit of zero roots $\{z_j \}$.
We note that Eq.\eqref{XXXE} is also
the energy of the Hamiltonian $\tilde{H}^{s+}$ \eqref{H1rational}.

The next task is to determine the values of roots $\{z_j \}$ that is to seek $2N+2$ constraints of roots $\{z_j \}$. By using the fusion technique, we obtain that the $\bar{\Lambda}(u)$ at the inhomogeneity points satisfies following functional relations
\bea
\bar{\Lambda}(\pm\theta_j)\bar{\Lambda}(\pm\theta_j-1)= \bar{d}^s (\pm\theta_j - 1)\bar{a}^s (\pm\theta_j),\quad j=1,\cdots N, \label{lam-lam}
\eea
where $\bar{d}^s (u) = \bar{a}^s (- u - 1)$ and $\bar{a}^s (u)$ is given by
\bea
\bar{a}^s (u) = \frac{u + 1}{u + \frac{1}{2}} (u + p) (\sqrt{1 + \xi^2} u +
   q) \prod_{j = 1}^N (u - \theta_j + 1) (u + \theta_j + 1).
\eea
From the direction calculation, we also know the values of $\bar{\Lambda}(u)$ at some special points,
\bea
\bar{\Lambda}(0)=\bar{\Lambda}(-1)=\bar{a}^s(0).
\eea
Thus the $2N + 2$ constrains of $\{z_j \}$ are
\bea
&&  4\prod_{l=1}^{2N+2} (\pm\theta_j - z_l i + \frac12) (\pm\theta_j - z_l i - \frac12) = \bar{a}^{s} (\pm\theta_j) \bar{d}^{s} (\pm\theta_j - 1), \, j = 1, \ldots, N, \label{lam-lam-a}\\
&&  2 \prod_{j = 1}^{2N + 2} \left( z_j i \pm\frac{1}{2} \right) =\bar{a}^s(0)  \label{lam-lam-a1},
\eea
which are the Bethe-ansatz-like equations satisfied by the zero roots $\{z_j \}$.

Similar with the anisotropic case, we consider that the inhomogeneity parameter are pure imaginary or zero. We denote $\theta_k=i \bar{\theta}_k$, where $\bar{\theta}_k$ is real. From the properties of $R$-matrix \eqref{Rxxx}, it is easy to prove that $\bar \Lambda(u)$ has the crossing symmetry
\bea
\bar{\Lambda}(u)=\bar{\Lambda}(-u-1).\label{lam-r}
\eea
Substituting \eqref{lambda} into \eqref{lam-r}, we have
\bea
\prod_{j = 1}^{2N + 2} \left( u - z_j i + \frac{1}{2} \right)=\prod_{j = 1}^{2N + 2} \left( u + z_j i + \frac{1}{2} \right).
\eea
Thus we conclude if $z_j$ is a zero root of $\bar{\Lambda}(u)$, then $-z_j$ must be the root.
One can also easily prove that
\bea\label{lambda-conj-r}
[\bar{t}(u)]^\dagger=\bar{t}(u^*), \quad [\bar{\Lambda}(u)]^*=\bar{\Lambda}(u^*).
\eea
Substituting \eqref{lambda} into \eqref{lambda-conj-r}, we have
\bea
\prod^{2N+2}_{k=1}\left(u^*+z_k^* i-\frac12\right)=\prod^{2N+2}_{k=1}\left(u^*+z_k i-\frac12\right),
\eea
which means that if $z_k$ is one zero root of $\bar{\Lambda}(u)$, then $-z_k$, $z_k^*$ and $-z_k^*$ must be the roots.

The distribution of zero roots with odd $N$ is slightly different from that with even $N$ at the ground state.
Because the odd and even $N$ give the same physical properties in the thermodynamic limit, we focus on even $N$.
We consider the case that $p>1/2$ and $q/\sqrt{1+\xi^2}<-1/2$.
From the numerical solutions of Bethe-ansatz-like equations with finite system sizes and the singularity analysis in the thermodynamic limit, we obtain that
the roots include the conjugate pairs $\{\bar{z}_j+i, \bar{z}_j-i\}$ and the additional roots $\pm z_a i$, where $\bar{z}_j$ and $z_a$ are real.
Because the spin exchanging interactions in Hamiltonian $\bar{H}^{s+}$ \eqref{H1_iso} are antiferromagnetic,
most zero roots form the bulk 2-strings rather than the real roots at the ground state.

Substituting the patterns of zero roots into the Bethe-ansatz-like equations \eqref{lam-lam-a}, taking the logarithm, and making the difference of the resulted equations for $\theta_j$ and $\theta_{j-1}$, in the thermodynamic limit, we have
\bea
&&N\int_{-\infty}^{\infty}[\bar{b}_3(\bar{\theta}-z)+\bar{b}_1(\bar{\theta}-z)] \bar{\rho}(z)dz +\bar{b}_{2z_a+1}(\bar{\theta})+\bar{b}_{2z_a-1}(\bar{\theta}) \no \\ &&= \bar{b}_{2}(\bar{\theta})- \bar{b}_{1}(\bar{\theta})+\bar{b}_{2p}(\bar{\theta})+ \bar{b}_{2q/\sqrt{1+\xi^2}}(\bar{\theta})+2N\int_{-\infty}^{\infty} \bar{b}_{2}(\bar{\theta}-\theta)\bar{\sigma}(\theta)d\theta,
\eea
where the function $\bar{b}_n(u)$ is defined as
\bea
\bar{b}_n(u)=[\ln (u+\frac{n i}2)+\ln(u-\frac{ni}2)]'=\frac{2u}{u^2+\frac{n^2}4}.
\eea
In the homogeneous limit, we set the density $\bar{\sigma}(\theta)$ of inhomogeneity parameters as the delta function, $\bar{\sigma}(\theta)=\delta(\theta)$.
Then, by using the Fourier transformation,
we obtain the density of zero roots as
\bea\label{XXXopenrho}
\tilde{\bar{\rho}}(k)&=&\frac{2\tilde{\bar{b}}_2(k)}{\tilde{\bar{b}}_1(k)+\tilde{\bar{b}}_3(k)} +\frac{1}{N[\tilde{\bar{b}}_1(k)+\tilde{\bar{b}}_3(k)]} [\tilde{\bar{b}}_2(k)-\tilde{\bar{b}}_1(k)+\tilde{\bar{b}}_{2p}(k) \no \\ [4pt] &&+\tilde{\bar{b}}_{2q/\sqrt{1+\xi^2}}(k) -\tilde{\bar{b}}_{2z_a+1}(k)-\tilde{\bar{b}}_{2z_a-1}(k)],
\eea
where
\bea
\tilde{\bar{b}}_n(k)=\int_{-\infty}^{\infty}\bar{b}_n(u)e^{-iku} du=-2\pi i \, \text{sign}(k)e^{-\frac{n|k|}2}.
\eea
Then the ground state energy of the Hamiltonian $\bar{H}^{s+}$ is
\bea
\bar{E}^{s+}_{g} &=& \frac{N}{4}\int_{-\infty}^{\infty}\left(e^{-\frac{3|k|}2} -e^{-\frac{|k|}2}\right) \tilde{\bar{\rho}}(k) d k +\frac{2}{1-4z_a^2} \no \\[4pt]
&=& N(1-2\ln2)+\frac{\pi - 1}2 - \ln 2 + \frac{1}{2| p |} +
\frac{\sqrt{1+\xi^2}}{2| q |} \no \\ [4pt] && - \int^{\infty}_0
\frac{e^{- | p | k} + e^{- | q | k/\sqrt{1+\xi^2}}}{1 + e^{- k}} d k.
\eea
We see that the additional roots do not have the contribution to the ground state energy.

The ground state energy of the $XXX$ spin chain with periodic boundary condition
is $N(1-2\ln2)$ \cite{wang15}. Thus the surface energy of the system \eqref{H1rational} or \eqref{H1_iso} is
\bea
\bar{E}^{s+}_{s} = \frac{\pi - 1}2 - \ln 2 + \frac{1}{2| p |} +
\frac{\sqrt{1+\xi^2}}{2| q |} -  \int^{\infty}_0
\frac{e^{- | p | w} + e^{- | q | w/\sqrt{1+\xi^2}}}{1 + e^{- w}} d w.
\eea
With the help of decomposition \eqref{apop1}, we obtain the surface energy of the isotropic $D^{(1)}_2$ spin chain \eqref{apop} as
\bea
\bar{E}_s = \bar{E}^{s+}_{s} + \bar{E}^{s+}_{s}|_{\{s, s_1, s_2, s', s'_1, s'_2\}\rightarrow \{t, t_1, t_2, t', t'_1, t'_2\}}.
\eea

\section{Discussion}

We have studied the exact solution of anisotropic $D^{(1)}_2$ quantum spin chain with generic non-diagonal boundary magnetic fields.
We find that the transfer matrix of the system can be constructed by two six-vertex models with boundary reflections.
Based on the factorization identities, we obtain the eigenvalues and the Bethe-ansatz-like equations of the model.
By using the $t-W$ scheme, we obtain the patterns of zero roots of the transfer matrix. Based on them, the thermodynamic limit, ground states energy and surface energy are obtained.
We also give the results for the isotropic couplings case.

Starting from the obtained eigenvalues \eqref{ta} and \eqref{1BA012}, the corresponding eigenstates can be retrieved by using the separation of variables for the integrable systems \cite{Skl95,Fra08, Fra11,Nic12} or the off-diagonal Bethe ansatz \cite{Zhan15,Zhan15-1}.
Then the correlation functions, norm, form factors and other interesting scalar products can be calculated.
Based on the patterns of zero roots, other physical quantities such as the elementary excitations and the free energy at finite temperature
can be studied.
We also note that the results given in this paper are the foundation to exactly solve the high rank $D^{(1)}_{n}$ model with the nested method.

\section*{Acknowledgments}

The financial supports from National Key R$\&$D Program of China
(Grant No. 2021YFA1402104), the National Natural Science
Foundation of China (Grant Nos. 12074410, 12047502, 12075177,
12147160, 11934015, 11975183 and 11947301), Major Basic Research
Program of Natural Science of Shaanxi Province (Grant Nos.
2021JCW-19, 2017ZDJC-32), Australian Research Council (Grant No.
DP 190101529), Strategic Priority Research Program of the Chinese
Academy of Sciences (Grant No. XDB33000000), Double First-Class
University Construction Project of Northwest University, and the
fellowship of China Postdoctoral Science Foundation (2020M680724)
are gratefully acknowledged.

\end{document}